\documentclass{article}
\usepackage[utf8]{inputenc}
\usepackage[T1]{fontenc}
\usepackage{amsmath}
\usepackage{graphicx}
\usepackage{psfrag}
\usepackage{listings}
\usepackage[usenames,dvipsnames,svgnames,table]{xcolor}
\usepackage{caption}
\usepackage{authblk}
\usepackage{verbatim}
\usepackage{lineno}
\usepackage[normalem]{ulem}

\usepackage[left=25mm,right=25mm,top=30mm,bottom=30mm]{geometry}

\usepackage{natbib}

\usepackage{hyperref}

\title{\bf Model of knowledge transfer within an organisation}
\date{\today}
\author{Agnieszka Kowalska-Stycze\'n\thanks{Agnieszka.Kowalska-Styczen@polsl.pl}} 
\affil{Silesian University of Technology,
Faculty of Organisation and Management,\\
ul. Roosevelta 26/28, 41-800 Zabrze, Poland}
\author{Krzysztof Malarz\thanks{Krzysztof.Malarz@fis.agh.edu.pl, WWW:~http://home.agh.edu.pl/malarz/}~}
\author{Kamil Paradowski}
\affil{{AGH University of Science and Technology},
{Faculty of Physics and Applied Computer Science},\\
al. Mickiewicza 30, 30-059 Krakow, Poland}

\date{\today}

\begin{document}
\maketitle

\begin{abstract}
Many studies show that the acquisition of knowledge is the key to build competitive advantage of companies.
We propose a  simple model of knowledge transfer  within the organization and  we   implement   the  proposed  model using cellular   automata   technique.
In this paper the organisation is considered in the context of complex systems.
In  this perspective, the main role in organisation is played by the network of informal contacts and the distributed leadership.
The goal of this paper is to check which factors influence the efficiency and effectiveness of knowledge transfer.
Our studies indicate a significant role of initial concentration of chunks of knowledge for knowledge transfer process,  and the results suggest taking action in the organisation to shorten the distance (social distance) between people with different levels of knowledge, or working out incentives to share knowledge.
\end{abstract}



\section{Introduction}

In today's rapidly changing environment, the knowledge is a dominant source of organisation sustainable competitive advantage~\citep{Chen-2004,Lyles-1996,Tsai-2001}, thus, the ability to acquire information and the knowledge creation is a big organisations force~\citep{Nonaka-1994,Zander-1995}.
Moreover,~\citet{Nonaka-1995} postulate that knowledge is no longer one of the traditional elements of production, but it becomes the only factor of production, which determines the company competitiveness.
Therefore, the knowledge transfer  is one of the most important elements in the management process, especially, for the organisational change~\citep{Martinez-2016}, the project management~\citep{Spalek-2014}, and widely understood development.
Organisations that can manage changes and adaptations effectively will not only survive, but thrive~\citep{Brown-2003}.  The organisational change is a complex process and includes alterations in technology, structures or systems, it may be defined as alterations of existing work routines and strategies that affect a whole organisation~\citep{Herold-2008}.
 
Both the organisational change and the development require the transfer of knowledge, which allows people to acquire knowledge, while the organisations can distribute it and properly use. However, there are substantial barriers which make that the transfer knowledge within the firm is difficult and complicated too~\citep{Szulanski-1996}, the efforts to share knowledge are often impeded by employees' tendencies to guard and selectively share information~\citep{Gilmour-2003}.
Moreover, the knowledge transfer needs to span different knowledge holders and requires a collaborative effort of both, knowledge recipients and senders.
A very important element for an efficient knowledge transfer is thus a senders' disseminative capacity and a receivers' absorptive capacity~\citep{Tang-2011}.
The disseminative capacity can be defined as the network members ability to effectively and efficiently communicate and transmit knowledge to other network members, whereas the absorptive capacity is the ability of the recipient to receive a knowledge. The knowledge can be explicit or tacit~\citep{Nonaka-1995}.
The explicit knowledge is an information that is stored in the form of documents or other media, while the tacit knowledge is an information that results from a person's experience, consists partly of technical skills~\citep{Novianto-2012}. Particularly, the study of transfer of tacit knowledge seems to be interesting, because this kind of knowledge is difficult to spread among members within an organisation~\citep{Tang-2011,Teece-2000,Tsai-2002}.

Many studies highlight also the importance of social interaction among organisational members in knowledge exchange~\citep{Chen-2007,Ibarra-1993,Tsai-2002}.
They suggested that organisational units can leverage knowledge resources through interacting with one another.
Furthermore, interactions between organisation members are mainly informal, but such informal relations play an increasing role and they are the main source of influence in organisations in the context of organisation learning, innovation and adaptation processes~\citep{Ibarra-1993}. Such a point of view indicates the emergence and `bottom-up' approach to knowledge transfer, because this approach is also postulated in the change process~\citep{Butcher-2000,Higgs-2005,Kempster-2014}. Moreover, in organisation, knowledge more often moves in a horizontal direction and it is informal process~\citep{Girdauskiene-2012} whereas formal sharing of knowledge can reach much broader populations but may stifle some of spontaneous and creative aspects of the informal sharing modes~\citep{ODeal-1999}. In this article, knowledge transfer is thus understood as a common process in a creative organisation, which is mostly implemented informally by sharing knowledge `face to face'~\citep{Girdauskiene-2012}.

It should be added, that in this knowledge transfer process, leaders also play an important role~\citep{Girdauskiene-2012}. \citet{Higgs-2003} points that leadership effectiveness is increasingly moving away from leader-centric and `top-down' approach, whereas particularly attractive is perspective of distributed leadership~\citep{Kempster-2014}. This kind of leadership is an `emergent property of a group or network of interacting individuals working with openness of boundaries [$\cdots$ and] the varieties of expertise are distributed across the many, not the few'~\citep[p. 7]{Bennet-2003}.
In other words, distributed leadership implies that, `the leadership function is stretched over the work of a number of individuals, and the task is accomplished through interaction and collective action'~\citep{Harris-2013}.

Despite burgeoning literature on knowledge transfer, relatively little is known about how strategic knowledge is created and exchanged.
As behaviours that come from interactions of individuals and groups are characterised by complexity, dynamics, adaptation and non-linearity, they are very difficult for empirical studies.
This means, than virtual simulation can be a useful tool to explore knowledge transfer dynamics.
We therefore propose a model of knowledge transfer based on the cellular automaton (CA)~\citep{Wolfram-2002,Ilachinski-2001}.

In CA technique the model system is represented as a regular grid of cells.
For each site $i$ a scalar variable $s^i$ is assigned from finite set of possible states $s^i\in\mathcal{S}$.
The rule $\mathcal{F}$ maps the state of $i$-th cell in time $t$ into state of this cell $i$ in the next time step $t+1$ basing on states of sites in $i$-th site neighbourhood $\mathcal{N}^i$
\[
s^i(t+1)=\mathcal{F}(\mathcal{N}^i(t)).
\]
For a square lattice the simplest neighbourhood $\mathcal{N}$ is von Neumann one, which for discrete coordinates consists site $(i,j)$ and its four topologically nearest neighbours at $(i-1,j)$, $(i+1,j)$, $(i,j-1)$, $(i,j+1)$.

The rule $\mathcal{F}$ is applied synchronously to all sites in the system.

In our approach we extend a classical definition of CA scalar variable $s^i$ to the vector one $\mathbf{C}^i$.
Formally our set of states $\mathcal{S}$ has $2^K$ elements in $\{ 0,1,\cdots,2^K-1 \}$, however, we prefer to think about binary representation of these numbers, i.e. on $K$-elements long Boolean vectors $s^i=[c_1^i,c_2^i,\cdots,c_K^i]_{\text{bin}}=\mathbf{C}^i$.
Such representation yields comfortable interpretation of $c_k^i$ in terms of our conceptual model described below, however, the rigorous mathematical and formal definition of transition function $\mathcal{F}$ become a little bit cumbersome---thus we will provide its definition in rather narrative manner.

The rest of this paper is organised as follows: The next section presents the model of knowledge transfer (concept of the model, formal model and design of virtual experiments). Subsequently, the results of our simulations are analysed. Then, discussion and conclusions are presented. Finally, we propose future research.

\section{\label{S:model}Model}

\subsection{\label{S:conceptmodel}Conceptual model}

To allow analysis of the dynamics of the transfer of knowledge in the perspective of complex systems, in this paper, a CA model has been proposed.
The basic assumption of the model is a division of transferred knowledge into a certain amount of chunks.
This concept was inspired by studies of \citet{Reagans-2003}.
The chunks of knowledge proposed here can be understood  in many ways.  
For example, in the perspective of knowledge management, they may be elements of tacit knowledge in organisational model of knowledge creation, proposed by \citet{Nonaka-1995}.
In the approach of cited authors, knowledge (and thus also its chunks) applies to both information and beliefs and expectations.

Transfer of knowledge is also a fundamental process of implementing the function of cognitive information culture~\citep{Zbiegien-1999}.
In this perspective, the proposed `chunks of knowledge' will be associated with providing a general knowledge and information about the world.
People come together to interact in the organisational context, acquire knowledge, broaden horizons, transmit to each other the interpretations of events and the comments.
They also show their evaluative opinions, which obviously can affect attitudes and provide ready-made patterns of behaviour (e.g. in relation to the events associated with organisational change).

The adoption of CA model with lattice fully populated by agents as a model organisation, allows for reflecting a dense network of social interaction, whereas, the von Neumann neighbourhood of four neighbours reflects the network, which takes account only strong ties among agents.
Such assumptions in simulation studies, do not allow admittedly (at this stage of the model development) to analyse the impact of social cohesion and range on the effectiveness of knowledge transfer but largely reflect the results of empirical studies of~\citet{Reagans-2003}.
Static regression models of the network parameters impact on the easiness of knowledge transfer, obtained by the authors in empirical studies suggest first of all that  strong ties and a dense network, facilitate the transfer.
Also,  the number of strong ties, declared by the cited authors---on average less than six---seems to be well approximated  by adopted von Neumann neighbourhood.

The adopted model rules are a compromise of different perspectives of the knowledge transfer mechanism.
First of all, it is assumed that the agents are members of the organisation/community and they have (at the beginning of the simulation), a certain amount of  the common knowledge (chunks of knowledge). This common knowledge is the result of  the organisation activities, for example,  in the framework of the knowledge management policy.
The model assumes that each agent with equal probability, can absorb each portion of knowledge `offered' by the organisation. The level of probability reflects the quality of the activities of the formal `knowledge leadership' (e.g., this is the result of investment in the area of knowledge management).

Secondly, in accordance with the concept of distributed leadership, each agent can be a leader or a follower.
The agent role in the step of interaction depends on the quantity and quality of owned chunks of knowledge.
In a single interaction, an agent who has more knowledge (more chunks) is the leader.

Thirdly, in accordance with empirical findings of \citet{Reagans-2003} the transfer of knowledge is effective, if the distance of common knowledge between the source and recipient is small.
The model simplification introduce blocking the transfer of knowledge between agents (the leader and follower) if a distance of their knowledge (measured by the difference in the number of chunks of knowledge) is larger than one.
This approach is not much different from the opinion exchange model of~\citet{Deffuant-2000} when interaction among agents is possible only when the distance between agents opinions in one- \citep{Hegselmann-2002,Malarz2006b,Zhao-2016,Dong-2016} or two-dimensional \citep{Kulakowski-2009,Kulakowski2014,Gronek2011} space of opinions is smaller than assumed confidence level.

\subsection{\label{S:computmodel}Computerised model}

The model organisation is a $L\times L$ large square lattice with helical boundary conditions and with von Neumann neighbourhoods, i.e. the nearest neighbours of $i$-th agent ($i=1,\cdots,L^2$) are agents occupying nodes at $i\pm 1$ and $i\pm L$. 
The agents occupying virtual positions $i=-L+1,\cdots,0$ and $i=L^2+1,\cdots,L^2+L$ mirror opposite edges of the system, i.e. they are doppelgangers of agents from sites $i=L^2-L+1,\cdots,L^2$ and $i=1,\cdots,L$, respectively.

Each agent may posses some of $K$ chunks of knowledge $\mathbf{C}^i=[c_1^i,c_2^i,\cdots,c_K^i]$ and $c_k^i\in\{0,1\}$, where $c_k^i=0$ ($c_k^i=1$) indicates that this particular $k$-th chunks of knowledge is absent (present) in the $i$-th agent's knowledge.
Initially (at $t=0$), each agent $i$ has each of these knowledge chunks $c^i_{k=1,\cdots,K}(t=0)=1$ with probability $p$.

Every time step agent $i$ may inherit {\em single} chunk of information from {\em one} of his/her neighbour.
The knowledge transfer is possible if in time $t$ the neighbour $n$ has {\em exactly one more} chunks of knowledge than agent $i$:
\[
c^i_k(t+1)=1 \iff c^i_k(t)=0, \quad c^n_k(t)=1, \quad \nu^n(t)=\nu^i(t)+1,
\]
where $\nu^n(t)$ and $\nu^i(t)$ are current numbers of chunks of knowledge for agents $n$ and $i$, respectively.
The chunk of knowledge which is absent in $\mathbf{C}^i$ but which is present in $\mathbf{C}^n$ (and thus may be transferred from agent $n$ to agent $i$) is selected randomly.
The automaton rule described above is applied simultaneously to all agents.
The Java applet allowing for system evolution observation is available in Appendix~\ref{java}.

\subsection{The design of experiment}

Experiments that have been conducted were designed to investigate what factors affect the effectiveness and efficiency of knowledge transfer,  because, as emphasized by~\citet{Perez-2008}, taking into account both dimensions of knowledge transfer, gives a complete picture of this transfer.
\citet[p. 663]{Daft-1998} defines effectiveness as `the degree to which goals are attaineds' and efficiency as `amount of resources used to produce a unit of output'.
In addition, according to~\cite{Perez-2008}, the comprehension and usefulness can be construed as reflecting knowledge transfer `effectiveness', while the speed and economy can be understood as reflecting 'efficiency' in the knowledge transfer process.
It should also be noted that, the level of adoption of each initiative by the recipient units, is one of the measure of the knowledge transfer effectiveness~\citep{Jensen-2007}, and the speed at which the receiver acquires the new insights and skills,  can be understood as reflecting `efficiency' in the knowledge transfer process~\citep{Perez-2008}.

In view of the above, as the dependent variable describing an effectiveness of knowledge transfer the following parameters were adopted: 
\begin{itemize}
\item maximal number of chunks of knowledge ($K$),
\item lattice size ($L$),
\item initial concentration of chunks of knowledge ($p$).
\end{itemize}
Additionally, as the dependent variable describing an efficiency of knowledge transfer, the time $\tau$ necessary for reaching a stationary state was adopted.

On the other hand, the following independent variables were chosen in the designed experiments:
\begin{itemize}
\item number of chunks of knowledge ($K$),
\item lattice size ($L$),
\item initial concentration of chunks of knowledge ($p$).
\end{itemize}

The following levels of parameters were assumed in the first study:
\begin{itemize}
\item $K=4$ (transfer of smaller amount of knowledge is required) and $K=8$ (transfer of greater amount of knowledge is required),
\item $L=5$ (the equivalent of a small organisation), $L = 20$ (the equivalent of an average organisation)
\item and $p\in\{0.2, 0.5, 0.8\}$.
\end{itemize}

As the dependent variables describing an effectiveness of knowledge transfer the following parameters were adopted:
\begin{itemize}
\item $n(k)$ --- the fraction of agents having $k$ chunks of knowledge,
\item $f(k)$ --- the fraction of agents having $k$-th chunk of knowledge $c_k$,
\item $\langle f\rangle$ --- the coverage of any chunks of knowledge $c_k$ in agents' knowledge, i.e. the fraction of knowledge chunks held by typical member of the organisation.
\end{itemize}

We performed also additional experiments (second-fourth experiments) to find the threshold values of the parameters described in the first study.

\section{\label{S:results}Results}

The results presented in Figs.~\ref{F:komp_vs_t_v2}-\ref{F:nkomp_vs_t_K8} and \ref{F:omni_vs_t}-\ref{F:fci_vs_Kp_v2} are obtained as a result of procedure of data aggregation from $M$ independent simulation with different initial distributions of chunks of knowledge $\mathbf{C}^i$ for all agents ($i=1,\cdots,L^2$).
The program allowing for reproduction of these results is attached in Appendix~\ref{f77} as Listing~\ref{lst:main}.
The only exception are results presented in Fig.~\ref{F:tau_vs_p} which were obtained using program listed in Listing~\ref{lst:tau}.
If the number of independent simulations $M$ is not explicit specified then $M=100$.

\subsection{\label{S:veryfication}Model verification}
The model verification for computer models `is defined as checking the adequacy among conceptual models and computerised models' \citep[p. 138]{David-2013}.

\begin{figure*}
\centering
\psfrag{t}{$t$}
\psfrag{n(k)}[][c]{$f(k)$ [\%]} 
\psfrag{n(ci)}[][c]{$f(k)$ [\%]}
\psfrag{c1}{$c_1$}
\psfrag{c2}{$c_2$}
\psfrag{c3}{$c_3$}
\psfrag{c4}{$c_4$}
\psfrag{c5}{$c_5$}
\psfrag{c6}{$c_6$}
\psfrag{c7}{$c_7$}
\psfrag{c8}{$c_8$}
\psfrag{von Neumann, L=5, K=8, p=0.200}{$L=5, K=8, p=0.2$}
\psfrag{von Neumann, L=5, K=8, p=0.500}{$L=5, K=8, p=0.5$}
\psfrag{von Neumann, L=5, K=8, p=0.800}{$L=5, K=8, p=0.8$}
\psfrag{von Neumann, L=20, K=8, p=0.200}{$L=20, K=8, p=0.2$}
\psfrag{von Neumann, L=20, K=8, p=0.500}{$L=20, K=8, p=0.5$}
\psfrag{von Neumann, L=20, K=8, p=0.800}{$L=20, K=8, p=0.8$}
\includegraphics[width=.450\textwidth]{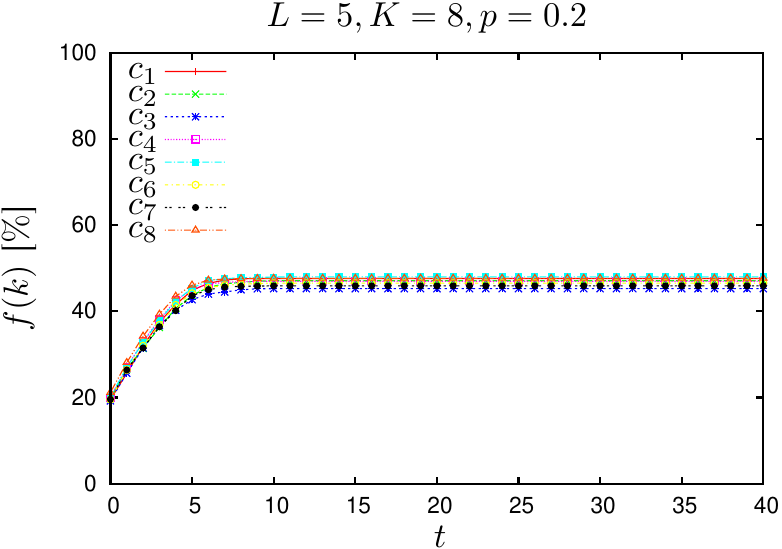}
\includegraphics[width=.450\textwidth]{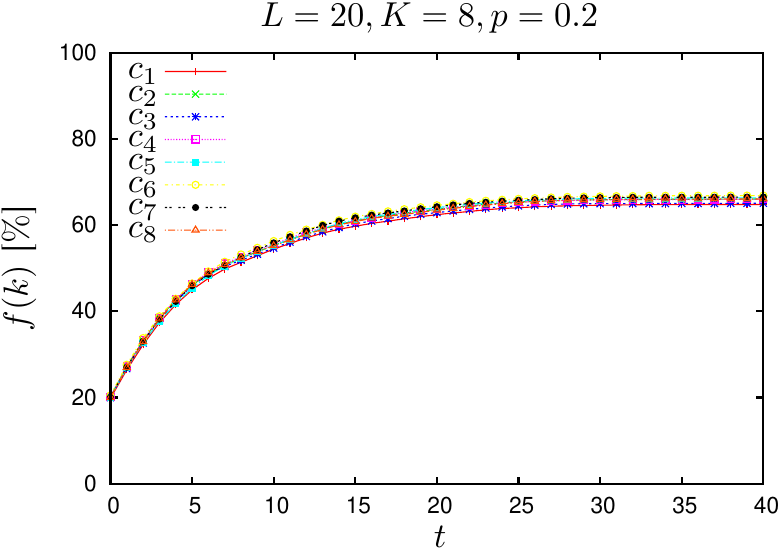}\\
\includegraphics[width=.450\textwidth]{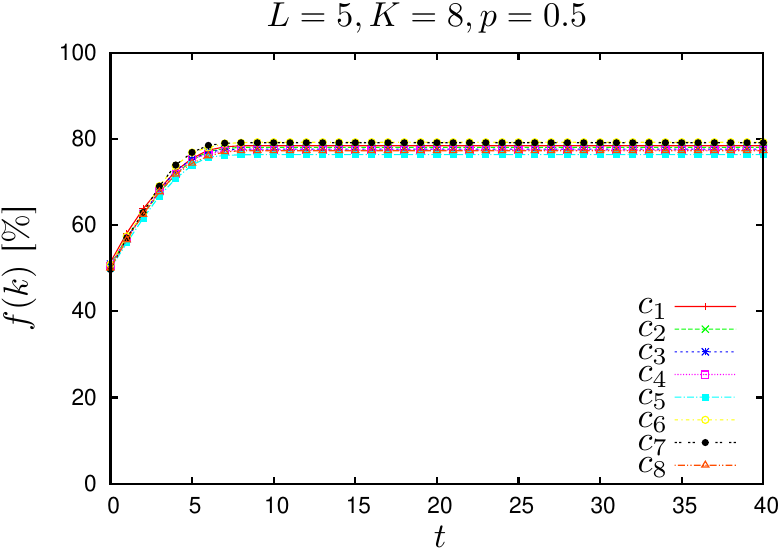}
\includegraphics[width=.450\textwidth]{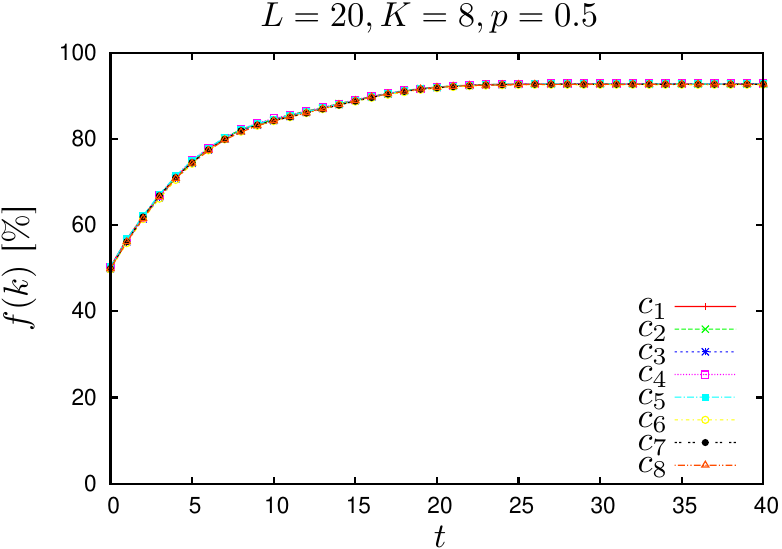}\\
\includegraphics[width=.450\textwidth]{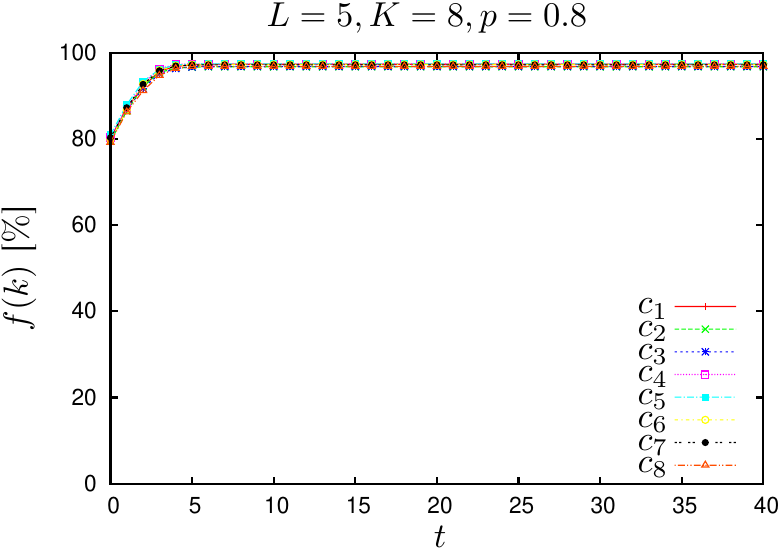}
\includegraphics[width=.450\textwidth]{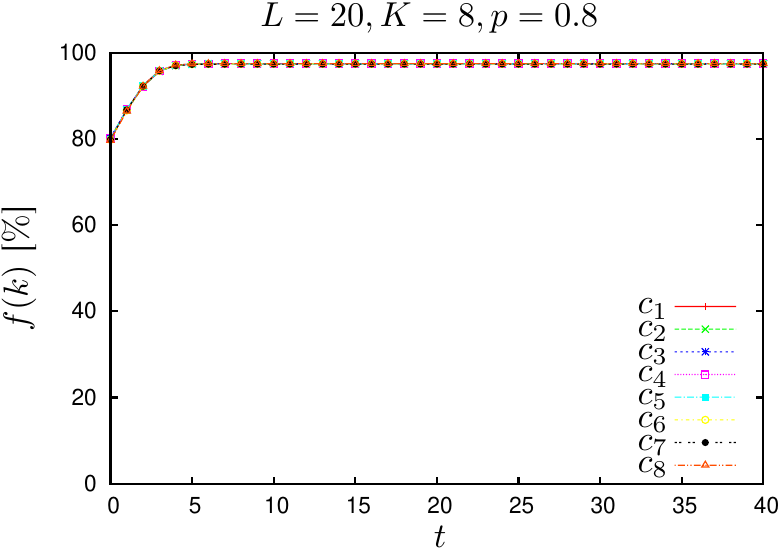}
\caption{\label{F:komp_vs_t_v2}The time evolution of the fraction $f(k)$ of agents having $k$-th chunk of knowledge $c_k$ for $K=8$ and various systems sizes ($L=5$, 20) and initial concentration of chunks of knowledge ($p=0.2$, 0.5, 0.8).
The values of $f(k)$ are averaged over $M=100$ independent simulations.}
\end{figure*}

In Listings~\ref{lst:exmples02} and~\ref{lst:exmples08} the values of $\mathbf{C}^i$ for small organisation ($L=5$) in several subsequent times steps $t<5$ are presented.
The vectors $\mathbf{C}^i$ between pairs of dashed lines correspond to $i=1,\cdots,5$ (in the first row), $i=6,\cdots,10$ (in the second row), etc.
The chunks of knowledge inherited in step $t$ are indicated by a red cross (x).
Eleventh lines of the listings contain information on global presence/absence of chunks of knowledge in organisation.
In both examples (Listings~\ref{lst:exmples02} and~\ref{lst:exmples08}) each chunk of knowledge is known at least by one of agents.
The lines 50--54 show time evolution of the aggregated number $F(k)\equiv\sum_{r=1}^M F_r(k)$ of agents having $k$-th chunk of knowledge $c_k$ and aggregated number $N(k)\equiv\sum_{r=1}^M N_r(k)$ of agents having exactly $k$ chunks of knowledge, where $F_r(k)$ [$N_r(k)$] are numbers of agents having $k$-th chunk of knowledge $c_k$ [numbers of agents having exactly $k$ chunks of knowledge] in $r$-th simulation.
These values (normalised to the system size $L^2$ and the number of simulations $M$) will be presented in Fig.~\ref{F:komp_vs_t_v2} 
\[
f(k)\equiv\dfrac{F(k)}{L^2}=\dfrac{\sum_{r=1}^M F_r(k)}{ML^2}
\]
and Figs.~\ref{F:nkomp_vs_t_K4}, \ref{F:nkomp_vs_t_K8} 
\[
n(k)\equiv\dfrac{N(k)}{L^2}=\dfrac{\sum_{r=1}^M N_r(k)}{ML^2}.
\] 

\lstset{morekeywords={x},keywordstyle=\color{red}\bfseries}
\begin{lstlisting}[float,frame=L,numbers=left,numberstyle=\tiny,basicstyle=\footnotesize,stepnumber=1,caption={Examples of model rules application for a small system $L=5$ and $K=4$ chunks of knowledge for $p=0.2$.},label=lst:exmples02]
 # L = 5, K = 4, M = 1, p = 0.200

 # t = 0
 --------------------------------------------------    i=  
[ 0 1 1 0][ 0 0 0 0][ 0 0 0 0][ 0 0 0 0][ 0 1 1 0]     01  02  03  04  05
[ 1 0 0 1][ 0 1 1 0][ 0 0 0 0][ 0 1 0 1][ 0 0 0 0]     06  07  08  09  10 
[ 0 0 1 0][ 1 1 0 0][ 0 0 0 0][ 0 0 1 0][ 0 0 0 0]     11  12  13  14  15
[ 0 0 0 0][ 0 0 1 1][ 0 0 0 0][ 0 0 1 0][ 0 0 1 0]     16  17  18  19  20
[ 1 0 0 0][ 1 0 0 1][ 0 0 1 0][ 0 1 1 0][ 0 1 0 1]     21  22  23  24  25
 --------------------------------------------------
 	global: [ 1 1 1 1] =  4

 # t = 1
 --------------------------------------------------
[ 0 1 1 0][ 0 0 0 0][ 0 0 x 0][ 0 0 0 0][ 0 1 1 0]
[ 1 0 0 1][ 0 1 1 0][ 0 0 0 0][ 0 1 0 1][ 0 0 x 0]
[ x 0 1 0][ 1 1 0 0][ 0 0 x 0][ 0 x 1 0][ 0 0 1 0]
[ 0 0 x 0][ 0 0 1 1][ 0 0 x 0][ 0 x 1 0][ 0 x 1 0]
[ 1 0 x 0][ 1 0 0 1][ 0 0 1 x][ 0 1 1 0][ 0 1 0 1]
 --------------------------------------------------

 # t = 2
 --------------------------------------------------
[ 0 1 1 0][ 0 0 x 0][ 0 0 1 1][ 0 0 x 0][ 0 1 1 0]
[ 1 0 0 1][ 0 1 1 0][ 0 0 x 0][ 0 1 0 1][ 0 x 1 0]
[ 1 0 1 0][ 1 1 0 0][ 0 x 1 0][ 0 1 1 0][ 0 x 1 0]
[ x 0 1 0][ 0 0 1 1][ 0 0 1 x][ 0 1 1 0][ 0 1 1 0]
[ 1 0 1 0][ 1 0 0 1][ 0 0 1 1][ 0 1 1 0][ 0 1 0 1]
 --------------------------------------------------

 # t = 3
 --------------------------------------------------
[ 0 1 1 0][ 0 0 1 x][ 0 0 1 1][ 0 x 1 0][ 0 1 1 0]
[ 1 0 0 1][ 0 1 1 0][ 0 0 1 x][ 0 1 0 1][ 0 1 1 0]
[ 1 0 1 0][ 1 1 0 0][ 0 1 1 0][ 0 1 1 0][ 0 1 1 0]
[ 1 0 1 0][ 0 0 1 1][ 0 0 1 1][ 0 1 1 0][ 0 1 1 0]
[ 1 0 1 0][ 1 0 0 1][ 0 0 1 1][ 0 1 1 0][ 0 1 0 1]
 --------------------------------------------------

 # t = 4
 --------------------------------------------------
[ 0 1 1 0][ 0 0 1 1][ 0 0 1 1][ 0 1 1 0][ 0 1 1 0]
[ 1 0 0 1][ 0 1 1 0][ 0 0 1 1][ 0 1 0 1][ 0 1 1 0]
[ 1 0 1 0][ 1 1 0 0][ 0 1 1 0][ 0 1 1 0][ 0 1 1 0]
[ 1 0 1 0][ 0 0 1 1][ 0 0 1 1][ 0 1 1 0][ 0 1 1 0]
[ 1 0 1 0][ 1 0 0 1][ 0 0 1 1][ 0 1 1 0][ 0 1 0 1]
 --------------------------------------------------
 
 # t |  F(1)  F(2)  F(3)  F(4) |  N(0)  N(1)  N(2)  N(3)  N(4) 
   0 |     4     7    10     5 |     9     6    10     0     0
   1 |     5    10    17     6 |     3     6    16     0     0
   2 |     6    13    20     8 |     0     3    22     0     0
   3 |     6    14    20    10 |     0     0    25     0     0
   4 |     6    14    20    10 |     0     0    25     0     0
\end{lstlisting}

\begin{lstlisting}[float,frame=L,numbers=left,numberstyle=\tiny,basicstyle=\footnotesize,stepnumber=1,caption={Examples of model rules application for a small system $L=5$ and $K=4$ chunks of knowledge for $p=0.8$.},label=lst:exmples08]
 # L = 5,  K = 4, M = 1, p = 0.800

 # t = 0
 --------------------------------------------------    i=  
[ 1 1 1 0][ 1 1 1 1][ 1 0 1 1][ 1 1 0 1][ 1 1 1 1]     01  02  03  04  05
[ 1 1 0 1][ 1 1 1 1][ 1 1 1 1][ 1 1 0 1][ 1 1 1 1]     06  07  08  09  10
[ 1 1 1 1][ 1 1 1 1][ 0 1 0 0][ 1 1 1 1][ 1 1 1 1]     11  12  13  14  15
[ 1 1 1 1][ 1 1 1 1][ 1 1 1 1][ 0 0 1 1][ 1 1 1 1]     16  17  18  19  20
[ 1 0 1 0][ 1 0 0 1][ 0 1 1 1][ 1 1 1 1][ 1 1 1 1]     21  22  23  24  25
 --------------------------------------------------
      global: [ 1 1 1 1] =  4

 # t = 1
 --------------------------------------------------
[ 1 1 1 1][ 1 1 1 1][ 1 x 1 1][ 1 1 x 1][ 1 1 1 1]
[ 1 1 x 1][ 1 1 1 1][ 1 1 1 1][ 1 1 x 1][ 1 1 1 1]
[ 1 1 1 1][ 1 1 1 1][ 0 1 0 0][ 1 1 1 1][ 1 1 1 1]
[ 1 1 1 1][ 1 1 1 1][ 1 1 1 1][ 0 0 1 1][ 1 1 1 1]
[ 1 x 1 0][ 1 x 0 1][ x 1 1 1][ 1 1 1 1][ 1 1 1 1]
 --------------------------------------------------

 # t = 2
 --------------------------------------------------
[ 1 1 1 1][ 1 1 1 1][ 1 1 1 1][ 1 1 1 1][ 1 1 1 1]
[ 1 1 1 1][ 1 1 1 1][ 1 1 1 1][ 1 1 1 1][ 1 1 1 1]
[ 1 1 1 1][ 1 1 1 1][ 0 1 0 0][ 1 1 1 1][ 1 1 1 1]
[ 1 1 1 1][ 1 1 1 1][ 1 1 1 1][ 0 0 1 1][ 1 1 1 1]
[ 1 1 1 1][ 1 1 1 1][ 1 1 1 1][ 1 1 1 1][ 1 1 1 1]
 --------------------------------------------------

 # t = 3
 --------------------------------------------------
[ 1 1 1 1][ 1 1 1 1][ 1 1 1 1][ 1 1 1 1][ 1 1 1 1]
[ 1 1 1 1][ 1 1 1 1][ 1 1 1 1][ 1 1 1 1][ 1 1 1 1]
[ 1 1 1 1][ 1 1 1 1][ 0 1 0 0][ 1 1 1 1][ 1 1 1 1]
[ 1 1 1 1][ 1 1 1 1][ 1 1 1 1][ 0 0 1 1][ 1 1 1 1]
[ 1 1 1 1][ 1 1 1 1][ 1 1 1 1][ 1 1 1 1][ 1 1 1 1]
 --------------------------------------------------

 # t = 4
 --------------------------------------------------
[ 1 1 1 1][ 1 1 1 1][ 1 1 1 1][ 1 1 1 1][ 1 1 1 1]
[ 1 1 1 1][ 1 1 1 1][ 1 1 1 1][ 1 1 1 1][ 1 1 1 1]
[ 1 1 1 1][ 1 1 1 1][ 0 1 0 0][ 1 1 1 1][ 1 1 1 1]
[ 1 1 1 1][ 1 1 1 1][ 1 1 1 1][ 0 0 1 1][ 1 1 1 1]
[ 1 1 1 1][ 1 1 1 1][ 1 1 1 1][ 1 1 1 1][ 1 1 1 1]
 --------------------------------------------------

 
 # t |  F(1)  F(2)  F(3)  F(4) |  N(0)  N(1)  N(2)  N(3)  N(4) 
   0 |    22    21    20    22 |     0     1     3     6    15
   1 |    23    24    23    23 |     0     1     1     2    21
   2 |    23    24    24    24 |     0     1     1     0    23
   3 |    23    24    24    24 |     0     1     1     0    23
   4 |    23    24    24    24 |     0     1     1     0    23
\end{lstlisting}

In Listing~\ref{lst:exmples02} the system evolution for $K=4$ and $p=0.2$ are presented.
During the first time step the agents (followers 3, 10, 11, 13, 14, 16, 18, 19, 20, 21 and 23) acquire new chunks of knowledge ($c_3$, $c_3$, $c_1$, $c_3$, $c_2$, $c_3$, $c_3$, $c_2$, $c_2$, $c_3$ and $c_4$, respectively).
These transfers come from local leaders (agents 23, 5, 6, 14, 9, 11, 19, 24, 25, 1, 22).
Please note that some agents (11, 14, 19) play  both roles and simultaneously send and receive chunks of knowledge.
The transfer of $c_3$ to agent 18 is possible from agents 19 and 23 as $c_3^{18}(t=0)=0$ and $\nu^{18}(t=0)=0$ and simultaneously $c_3^{19}(t=0)=1, \nu^{19}(t=0)=1$ and $c_3^{23}(t=0)=1, \nu^{23}(t=0)=1$.
Agent 18 chooses agent from whom he/she inherit $c_3$ randomly.

In Listing~\ref{lst:exmples08} the system evolution for the same set of parameters as in Listing~\ref{lst:exmples02} is presented.
The only exception is that the value of $p$ is four times larger ($p=0.8$ instead of 0.2).
The larger value of $p$ results in higher coverage of chunks of knowledge $\{F(c_1),\cdots,F(c_4)\}=\{23,24,24,24\}$ ($p=0.8$) instead of $\{ 6,14,20,10\}$ for $p=0.2$.
Also qualitative difference in the number of agents $N(k)$ possessing $k$ chunks of  knowledge is observed as for $p=0.2$ agents with $k=K=4$ chunks of knowledge are absent while they are dominant (23/25) fraction of $N(k)$ distribution for $p=0.8$.

Please note that although in both cases all chunks of knowledge are available in the system the evolution stops before diffusion of all chunks of knowledge to all agents.

As both, the initial distributions of chunks of knowledge $c_k$ among agents and the missing chunks of knowledge which will be inherited are selected randomly we do not expect any difference in $f(k)$ time evolution.
And indeed, curves of $f(k)$ vs. $t$ for various $k$ collapse into single curve for fixed values of $L$, $K$ and $p$ as presented in Fig.~\ref{F:komp_vs_t_v2}. 

\subsection{\label{S:main}Main results}

\subsubsection{\label{S:E1}First experiment: $n(k)$}

\begin{figure*}
\centering
\psfrag{t}{$t$}
\psfrag{k}{$k=$}
\psfrag{n(k)}[][c]{$n(k)$ [\%]}
\psfrag{von Neumann, L=5, K=4, p=0.200}{$L=5, K=4, p=0.2$}
\psfrag{von Neumann, L=5, K=4, p=0.500}{$L=5, K=4, p=0.5$}
\psfrag{von Neumann, L=5, K=4, p=0.800}{$L=5, K=4, p=0.8$}
\psfrag{von Neumann, L=20, K=4, p=0.200}{$L=20, K=4, p=0.2$}
\psfrag{von Neumann, L=20, K=4, p=0.500}{$L=20, K=4, p=0.5$}
\psfrag{von Neumann, L=20, K=4, p=0.800}{$L=20, K=4, p=0.8$}
\includegraphics[width=.450\textwidth]{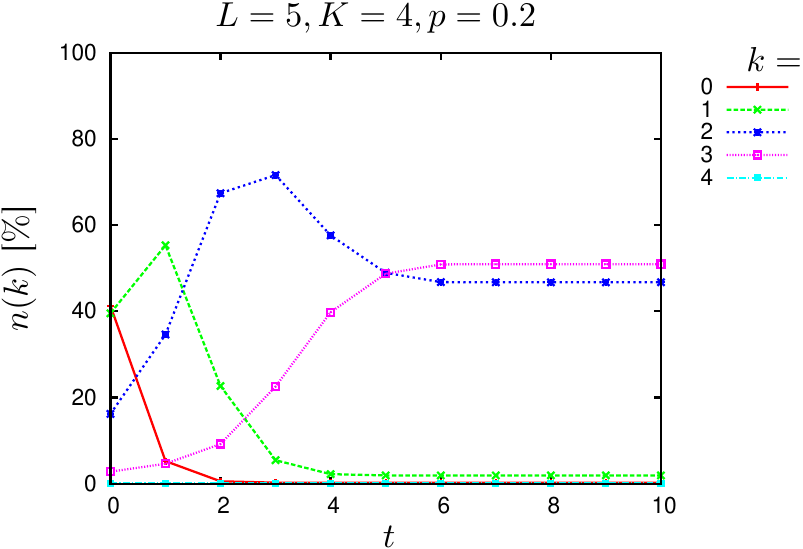}
\includegraphics[width=.450\textwidth]{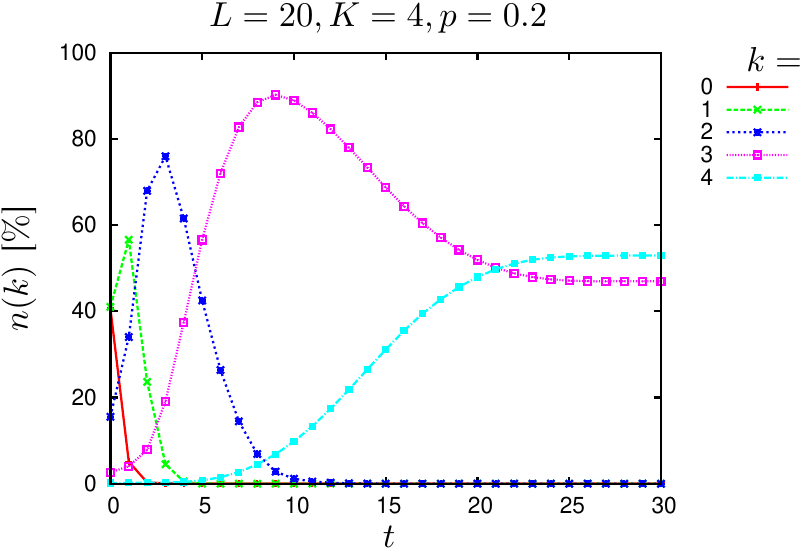}\\
\includegraphics[width=.450\textwidth]{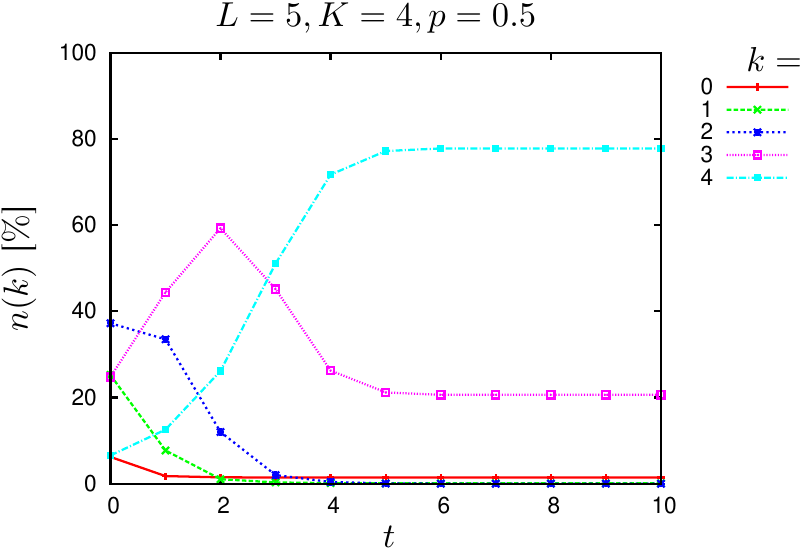}
\includegraphics[width=.450\textwidth]{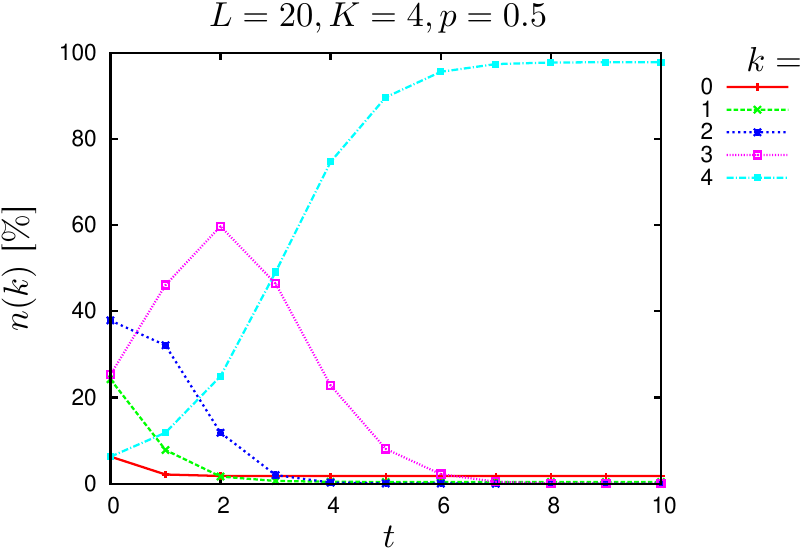}\\
\includegraphics[width=.450\textwidth]{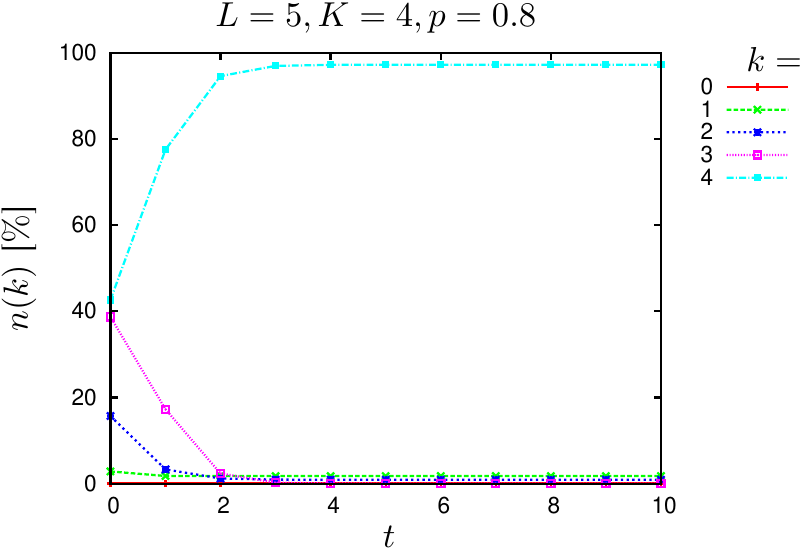}
\includegraphics[width=.450\textwidth]{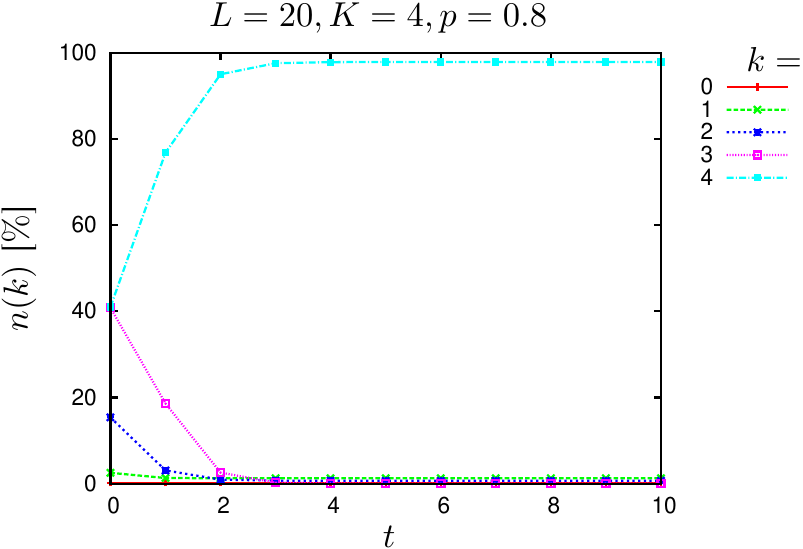}
\caption{\label{F:nkomp_vs_t_K4}The time evolution of the fraction $n(k)$ of agents having $k$ chunks of knowledge for $K=4$ and various systems sizes ($L=5$, 20) and initial concentration of chunks of knowledge ($p=0.2$, 0.5, 0.8).
The values of $n(k)$ are averaged over $M=100$ independent simulations.}
\end{figure*}

\begin{figure*}
\centering
\psfrag{t}{$t$}
\psfrag{k}{$k=$}
\psfrag{n(k)}[][c]{$n(k)$ [\%]}
\psfrag{von Neumann, L=5, K=8, p=0.200}{$L=5, K=8, p=0.2$}
\psfrag{von Neumann, L=5, K=8, p=0.500}{$L=5, K=8, p=0.5$}
\psfrag{von Neumann, L=5, K=8, p=0.800}{$L=5, K=8, p=0.8$}
\psfrag{von Neumann, L=20, K=8, p=0.200}{$L=20, K=8, p=0.2$}
\psfrag{von Neumann, L=20, K=8, p=0.500}{$L=20, K=8, p=0.5$}
\psfrag{von Neumann, L=20, K=8, p=0.800}{$L=20, K=8, p=0.8$}
\includegraphics[width=.450\textwidth]{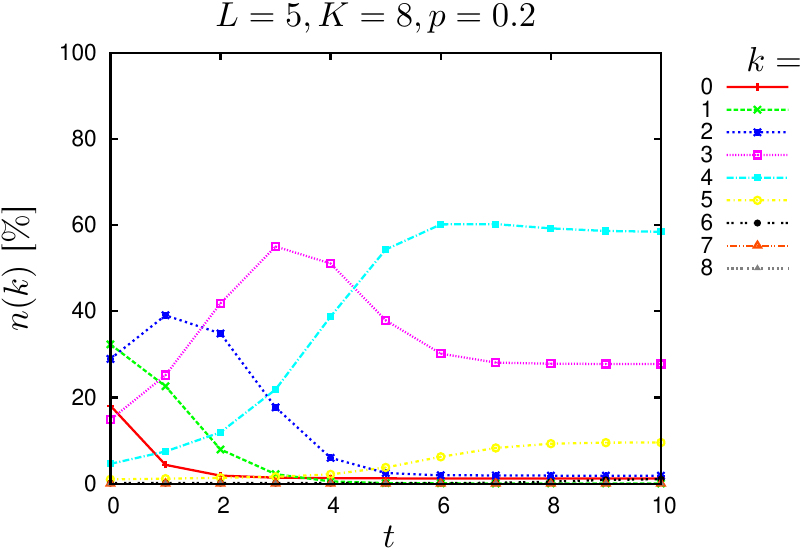}
\includegraphics[width=.450\textwidth]{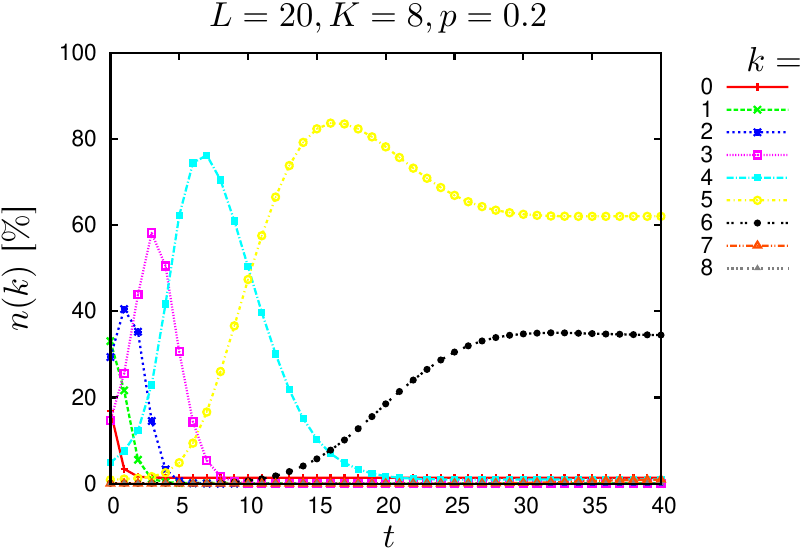}\\
\includegraphics[width=.450\textwidth]{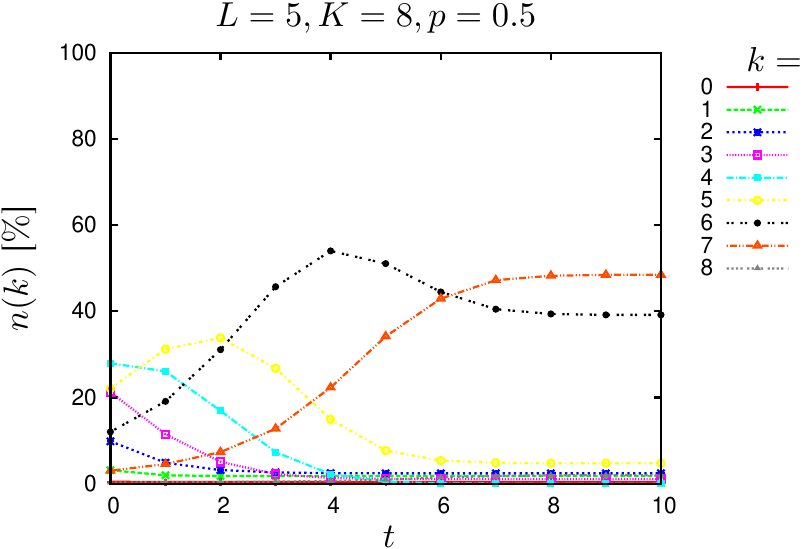}
\includegraphics[width=.450\textwidth]{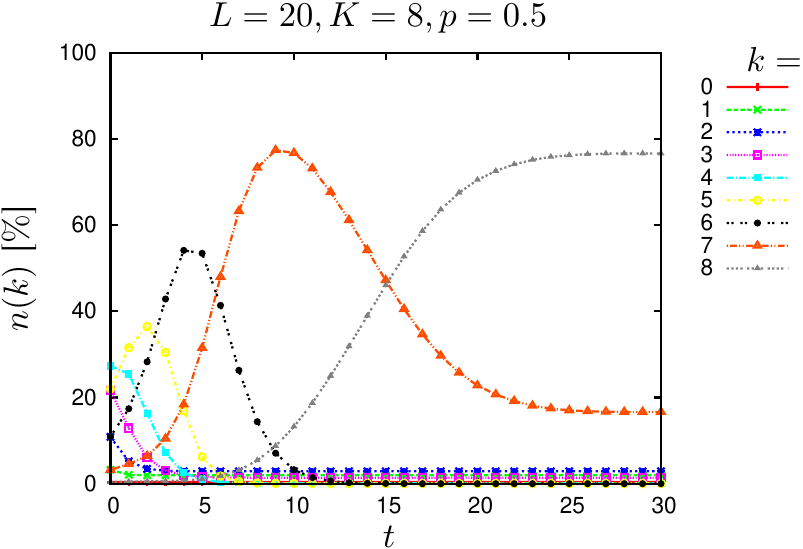}\\
\includegraphics[width=.450\textwidth]{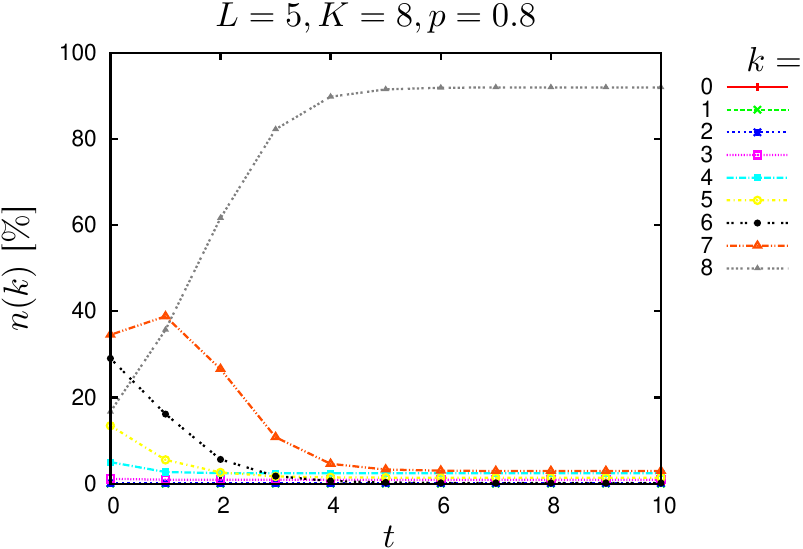}
\includegraphics[width=.450\textwidth]{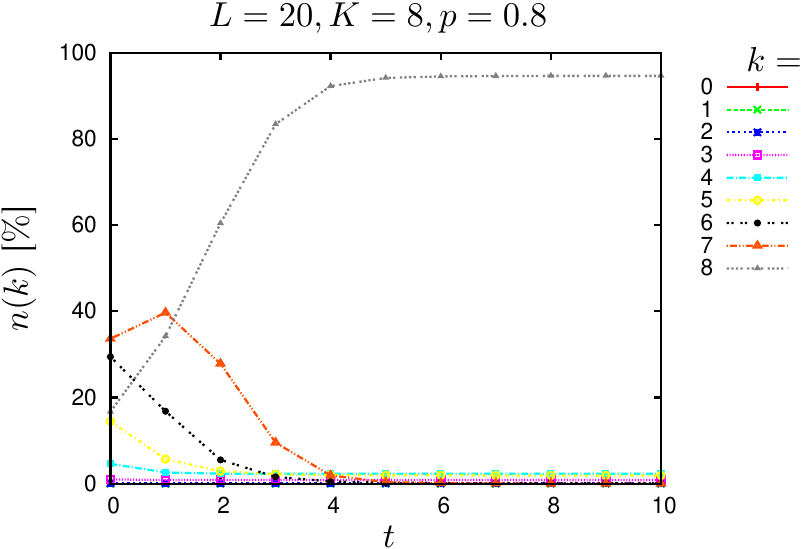}
\caption{\label{F:nkomp_vs_t_K8} The time evolution of the fraction $n(k)$ of agents having $k$ chunks of knowledge for $K=8$ and various systems sizes ($L=5$, 20) and initial concentration of chunks of knowledge ($p=0.2$, 0.5, 0.8).
The values of $n(k)$ are averaged over $M=100$ independent simulations.}
\end{figure*}

Initially, the evolution of agents fraction $n(k)$ who have $k$ required knowledge chunks depending on the simulation time $t$ was examined.
The study was conducted according to the experiment design for $K = 4$ (when organisation change requires less knowledge, Fig.~\ref{F:nkomp_vs_t_K4}) and for $K = 8$ (when organisation change requires more knowledge, Fig.~\ref{F:nkomp_vs_t_K8}).

In both cases, the simulations were conducted for a small organisation ($L=5$), and the average organisation ($L=20$) and three levels of initial concentration chunks of knowledge $p=0.2$, 0.5, 0.8.
As shown in Figs.~\ref{F:nkomp_vs_t_K4} and \ref{F:nkomp_vs_t_K8} the effectiveness of the knowledge transfer as measured by $n(k)$ depends on $L$ and $p$. 
The greater $p$ the greater the percentage of agents having a greater number of $k$ portions of all the required chunks of knowledge. 
For example, let us look at Fig.~\ref{F:nkomp_vs_t_K4} (left panel, for $L=5$ and $K=4$): for $p = 0.2$ the lack of agents having all four required knowledge chunks is observed, whereas for $p=0.5$, nearly 80\% of agents have four knowledge chunks and for $p=0.8$ almost 100\% of the agents have all required knowledge chunks.
In addition, to get a high enough percentage of the agents having $K$ portions of knowledge, the value of $p$ must be much greater for $K = 8$ (Fig.~\ref{F:nkomp_vs_t_K8}) than for $K=4$ (Fig.~\ref{F:nkomp_vs_t_K4}). 
For a larger system ($L=20$), the level of $p=0.2$ is high enough to make agents with all required chunks of knowledge the dominant fraction of agents in the organisation (for $p=0.2$, more than 50\% of agents have $k=4$ of the knowledge chunks, while agents with $k\le 2$ chunks of knowledge are absent). 
A similar relationship may be observed in Fig.~\ref{F:nkomp_vs_t_K8}, that is, in larger organisations, smaller value of $p$ is needed  to obtain a greater percentage of agents having the $k$ knowledge  chunks.

\begin{figure}[!ht]
\centering
\psfrag{L=5, M=1000}[][c]{(a) $L=5$}
\psfrag{L=20, M=1000}[][c]{(b) $L=20$}
\psfrag{p}{$p$}
\psfrag{tau}{$\tau$}
\psfrag{K=}{{\small $K=$}}
\includegraphics[width=.45\textwidth]{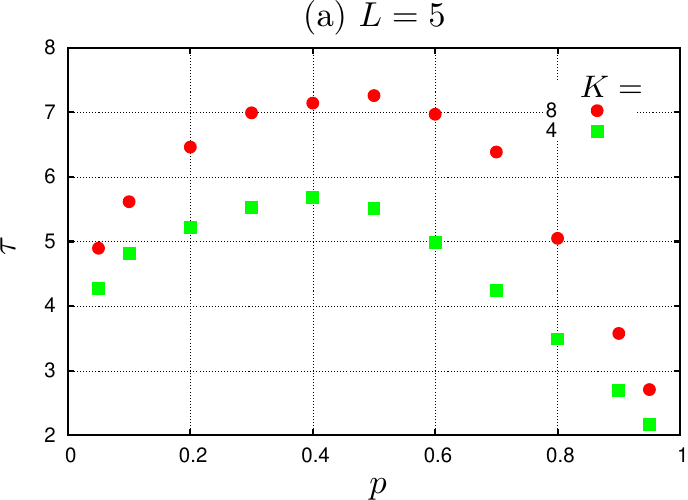}
\includegraphics[width=.45\textwidth]{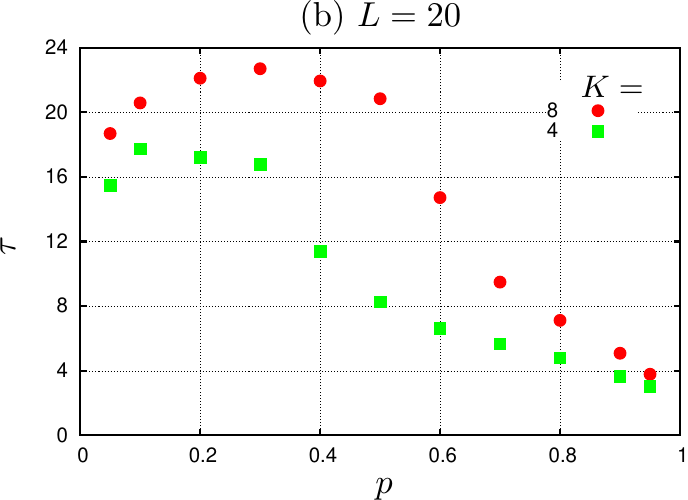}
\caption{\label{F:tau_vs_p}The time $\tau$ necessary for reaching the stationary state as dependent on initial concentration of chunks of knowledge $p$.
The results are averaged over $M=10^3$ independent runnings.
The error bars are smaller than the symbols.}
\end{figure}

The efficiency of the knowledge transfer may be quantitatively described as a time $\tau_r$ necessary for reaching the stationary state of the system during the $r$-th running.
The value of $\tau_r$ indicates the number of simulation time steps necessary for reaching the time point after which neither $F_r(k)$ ($k=1,\cdots,K$) nor $N_r(k)$ ($k=0,\cdots,K$) change any more.
The dependence of the average time $\tau=M^{-1}\sum_{r=1}^M\tau_r$ for small ($L=5$) and average ($L=20$) organisation sizes as dependent on initial fraction of chunks of knowledge $p$ for $K=4$ and $K=8$ are presented in Fig.~\ref{F:tau_vs_p}.
The results are averaged over $M=10^3$ independent simulations.
The obtained values of uncertainties of $\tau$, i.e. 
\begin{equation}
\label{eq:u}
u(\tau)=\sqrt{\dfrac{\sum_{r=1}^M (\tau_r-\tau)^2}{M(M-1)}}
\end{equation}
are smaller than symbol size.
In the case of a small organisation ($L=5$), the knowledge transfer time is much shorter than for the larger organisation ($L=20$).
Furthermore, this time is shorter for a smaller number of knowledge chunks (i.e., $K=4$) than for a larger one ($K=8$).
It may be also noted, that the knowledge transfer is most efficient for large values of $p$  ($p> 0.7$) when the small number of simulation steps are needed to achieve steady state.
In this case, i.e. for $p>0.7$, the difference between the times $\tau$ for $L = 5$ and $L = 20$ and $K = 4$ and $K = 8$ is the smallest.

\subsubsection{\label{S:E2}Second experiment: $n(K)$}
\begin{figure*}
\centering
\psfrag{von Neumann, L=5, K=2}[][c]{$L=5, K=2$}
\psfrag{von Neumann, L=20, K=2}[][c]{$L=20, K=2$}
\psfrag{von Neumann, L=5, K=3}[][c]{$L=5, K=3$}
\psfrag{von Neumann, L=20, K=3}[][c]{$L=20, K=3$}
\psfrag{von Neumann, L=5, K=4}[][c]{$L=5, K=4$}
\psfrag{von Neumann, L=20, K=4}[][c]{$L=20, K=4$}
\psfrag{von Neumann, L=5, K=8}[][c]{$L=5, K=8$}
\psfrag{von Neumann, L=20, K=8}[][c]{$L=20, K=8$}
\psfrag{p=}{$p=$}
\psfrag{t}{$t$}
\psfrag{omni}[][c]{$n(K)$ [\%]}
\includegraphics[width=.40\textwidth]{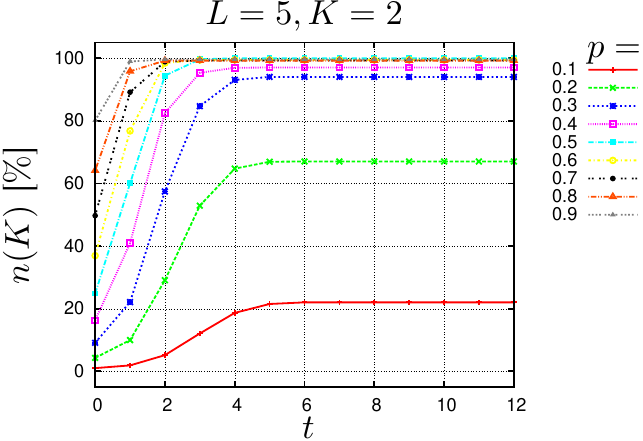}\hfill
\includegraphics[width=.40\textwidth]{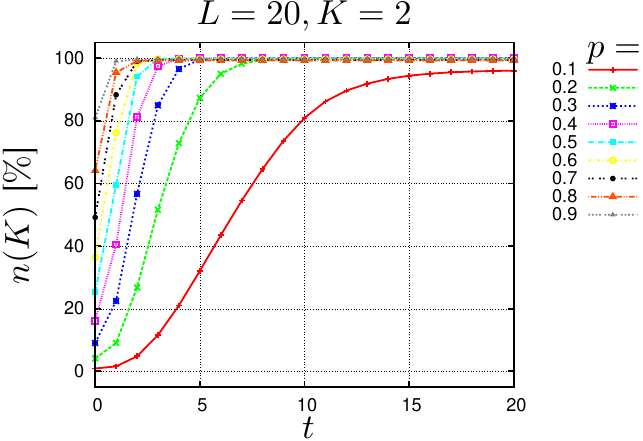}\\
\includegraphics[width=.40\textwidth]{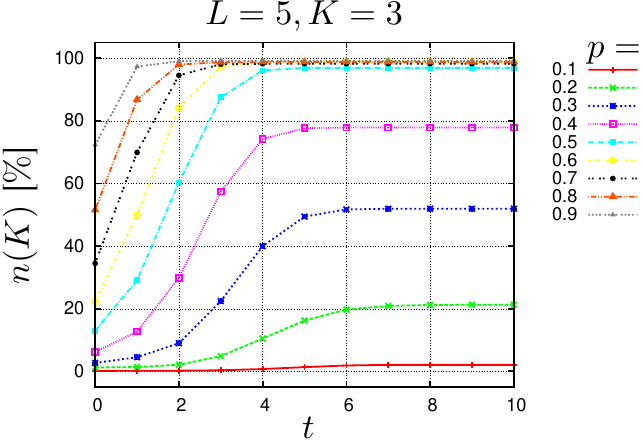}\hfill
\includegraphics[width=.40\textwidth]{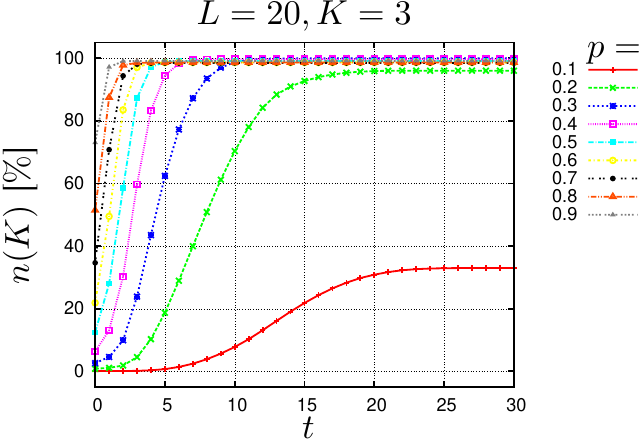}\\
\includegraphics[width=.40\textwidth]{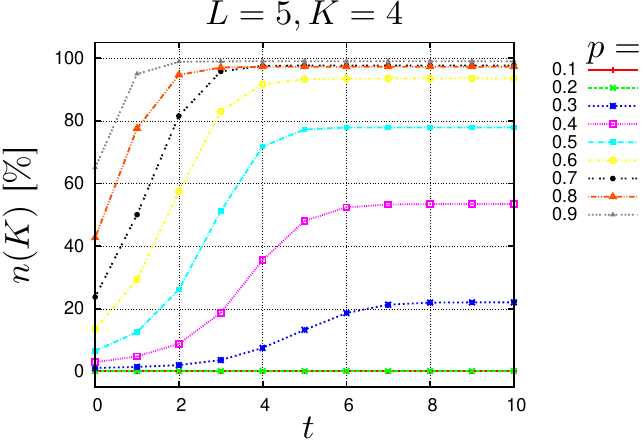}\hfill
\includegraphics[width=.40\textwidth]{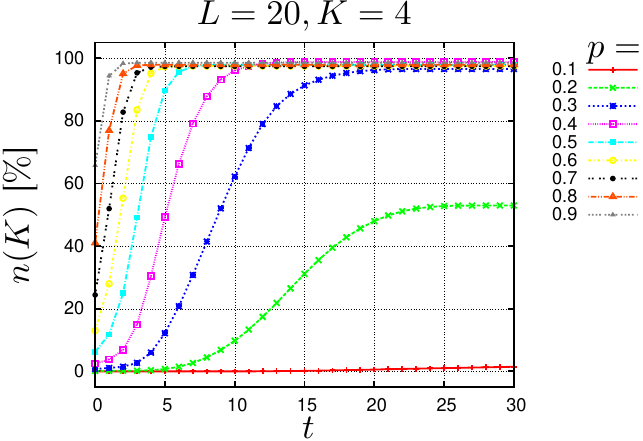}\\
\includegraphics[width=.40\textwidth]{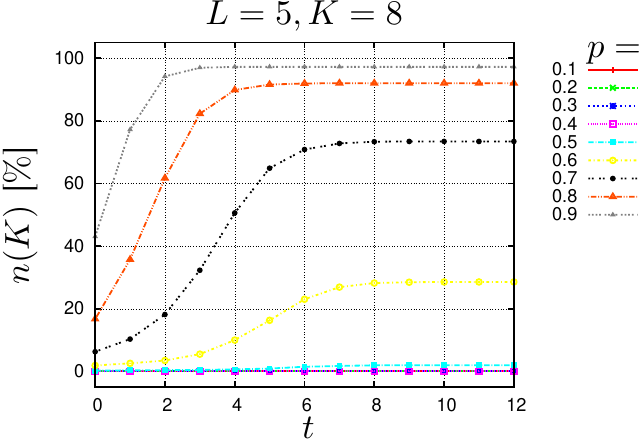}\hfill
\includegraphics[width=.40\textwidth]{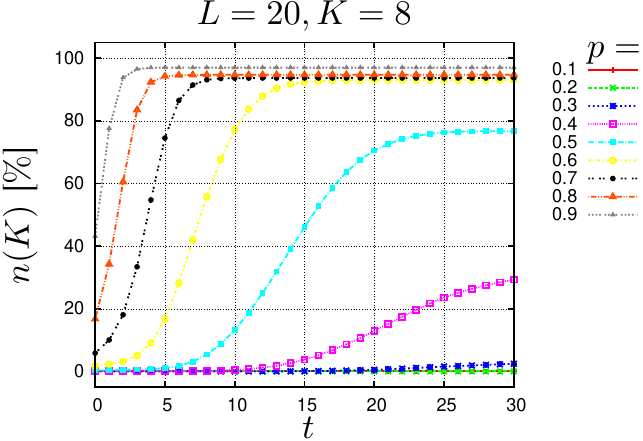}
\caption{\label{F:omni_vs_t}The time evolution of the fraction $n(K)$ of agents having total knowledge---i.e. possessing all $K$ chunks of knowledge $(c_1,c_2,\cdots,c_K)=(1,1,\cdots,1$).
The values of $n(K)$ are averaged over $M=100$ independent simulations.}
\end{figure*}

In Fig.~\ref{F:omni_vs_t} the results of the second experiment are presented.
Identical levels of $L$ (5 and 20) as in the first experiment have been adopted, whereas levels of $K$ (2, 3, 4, 8)  and $p$ (from 0.1 to 0.9 in steps of 0.1) have been expanded.
The basic result of conducted simulation confirmed the obtained dependencies  in the first study.
For larger lattice relatively lower value of $p$ is sufficient to achieve desired fraction $n(K)$ of agents having a complete knowledge. 
Furthermore, as can be seen in Fig.~\ref{F:omni_vs_t}, the greater the number of knowledge chunks $K$ is needed in organisation the greater must be initial fraction $p$ of knowledge chunks in order to make almost all the organisation members fully comprehensive, i.e. having all $K$ desired chunks of knowledge.
Please note however, that for some sets of parameters the full coverage of chunks of knowledge by all agents becomes impossible. 
For small organisation ($L=5$), $K=8$ and $p\ge 0.8$ the fraction of agents with total knowledge reaches 90\%$\le n(K)\le$100\%.
For average size of organisation ($L=20$) such level of agents knowledge is reached for $p\ge 0.6$.
On the other hand for $K=8$ and $p\le 0.5$ (small organisation) and for $p\le 0.3$ (average organisation) less than 10\% of agents posses all chunks of knowledge.

As can be seen in Fig.~\ref{F:omni_vs_t}, the efficiency of knowledge transfer is greater for smaller organisations, i.e. the stationary state is reached earlier for $L=5$ than for $L=20$.

\subsubsection{\label{S:E3}Third experiment: $n(K)$ vs. $L$ and $p$}
\begin{figure*}[!ht]
\centering
\psfrag{p}{$p$}
\psfrag{L, Nrun}{$L, M$}
\psfrag{(a) K=4, T=200}[][c]{(a) $K=4$}
\psfrag{(b) K=8, T=200}[][c]{(b) $K=8$}
\psfrag{(c)}[][c]{(c) $K=4$}
\psfrag{(d)}[][c]{(d) $K=8$}
\psfrag{\% omni(t->infty)}[][c]{$n(K)$ [\%]}
\includegraphics[width=.450\textwidth]{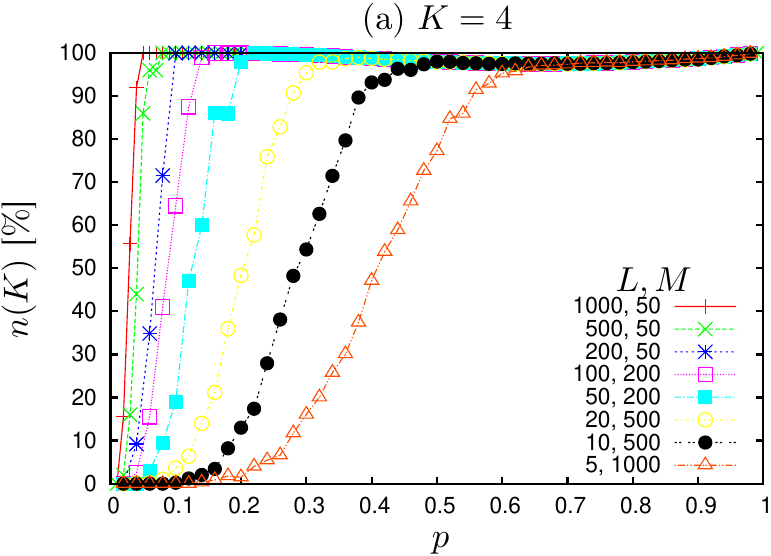}
\includegraphics[width=.450\textwidth]{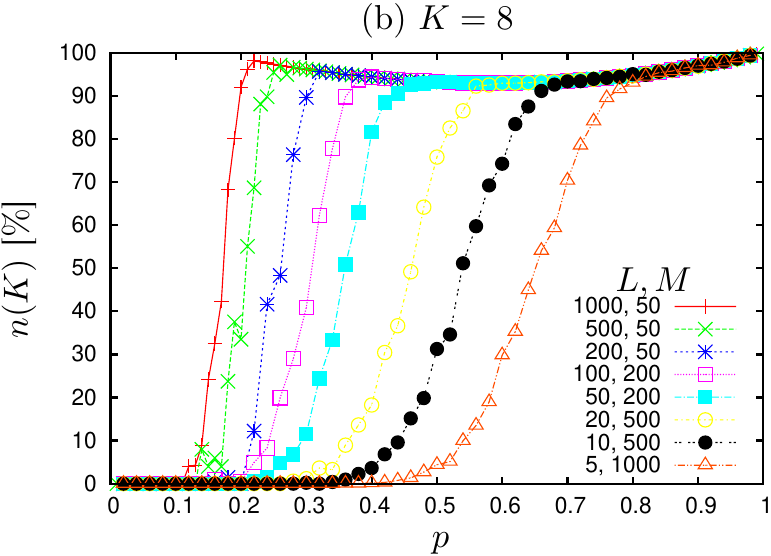}\\
\includegraphics[width=.450\textwidth]{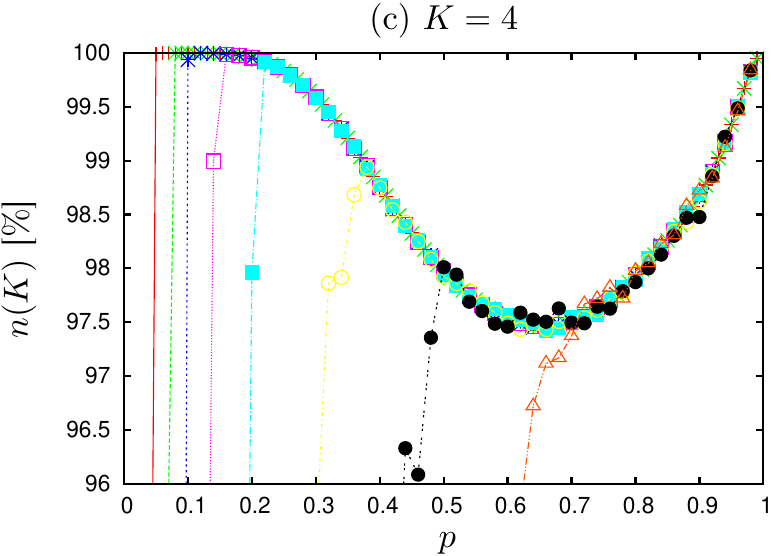}
\includegraphics[width=.450\textwidth]{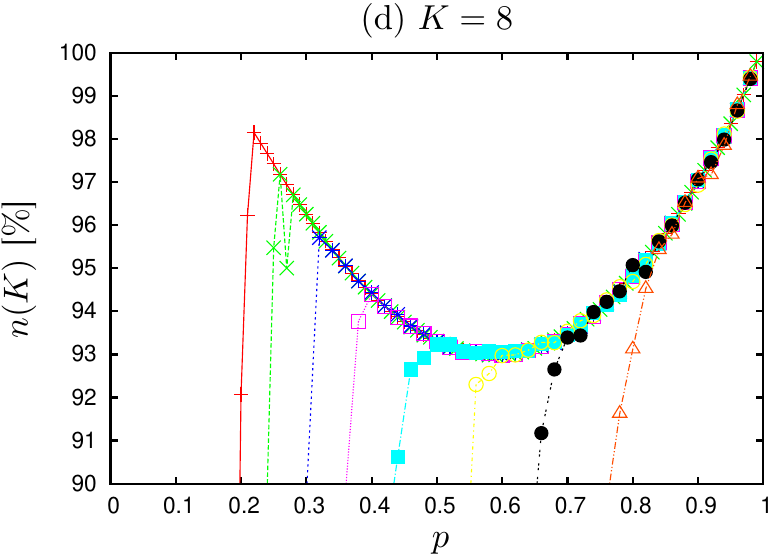}
\caption{\label{F:omniatT_vs_p}The fraction of agents having total knowledge ($c_1,c_2,\cdots,c_K)=(1,1,\cdots,1$) vs. initial probability $p$ for various system sizes $L$ and various sizes of total knowledge $K$.
The results are averaged over $M$ independent simulations.
The system reaches the stationary state in $t<200$ ($t<1000$) steps for $L<400$ ($L>400$).
In sub-figures (c) and (d) the upper part of sub-figures (a) and (b) are displayed ($90\%<n(K)<100\%$).}
\end{figure*}

\begin{figure}[!hb]
\centering
\psfrag{pC}{$p_C$}
\psfrag{pO}{$p_O$}
\psfrag{L}{$L$}
\psfrag{K=}{$K=$}
\psfrag{T=200}[][c]{(a)}
\psfrag{(b) T=200}[][c]{(b)}
\includegraphics[width=.450\textwidth]{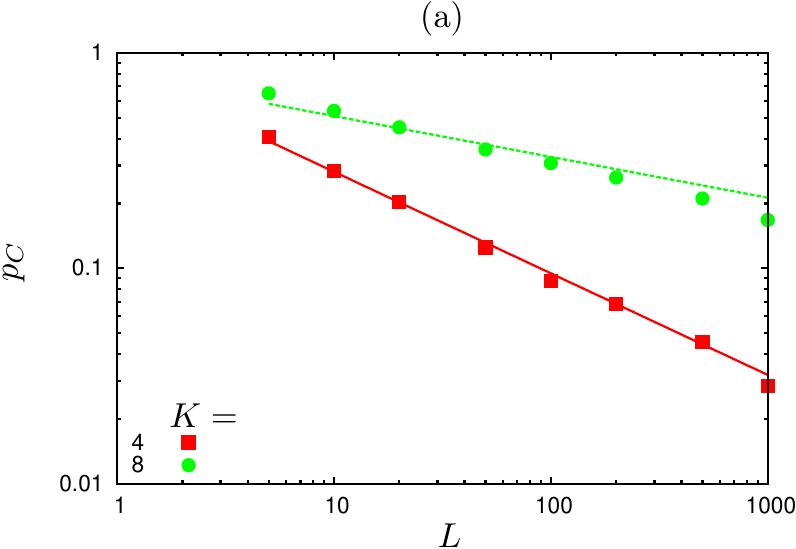}
\includegraphics[width=.450\textwidth]{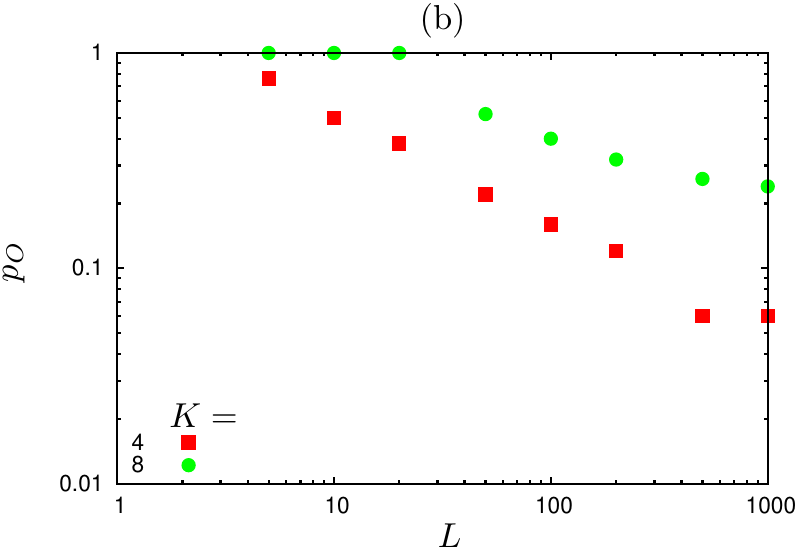}
\caption{\label{F:pCO_vs_L} (a) The critical values of initial concentration of chunks of knowledge $p_C$ as dependent on linear system size $L$.
For $p=p_C$ half of agents posses total knowledge.
(b) The optimal value of initial concentration of chunks of knowledge $p_O$ which guarantee the most efficient transfer of knowledge, i.e. giving the largest fraction of agents with $K$ chunks of knowledge as dependent on linear system size $L$.
For $p=p_O$ the dependencies presented in Fig.~\ref{F:omniatT_vs_p} have local maximum.
$T=200\, (L<400)$, $T=1000\, (L>400)$.}
\end{figure}

The third experiment was designed to precisely analyse the impact of the lattice size $L$ on the investigated dependent variables.
The experiments were conducted for $K = 4$ and $K=8$ (small and medium-sized organisations) and for the $p$ levels---as in the second experiment.
The size of the lattice (organisation) was changed to the levels $L=5$, 10, 20, 50, 100, 200, 500 and 1000.
The results of average values of $n(K)$ (independent replicates obtained for the simulation from 50 to 1000 runs) is shown in Fig.~\ref{F:omniatT_vs_p}.
The results are averaged over $M$ independent simulations as indicated in the picture inset.
The system reached the stationary state in $t<200$ ($t<1000$) steps for $L<400$ ($L>400$).
The bottom part of the figure shows enlarged fragments of the top figure part.
Dependencies shown in the figures, generally confirm the model behaviour obtained in the previous two studies.
First of all, the large size $L^2$ of modelled organisations require smaller values probability $p$ for a full (or nearly full, $n(K)\ge90\%$) knowledge into the system.

Let $p_C$ be an intitail fraction $p$ of chunks of knowledge which guarantee that half of agents will aquire the total knowledge.
The quantity $p_C$ may be treated as characteristic of efficiency of knowledge transfer complementary to time $\tau$.
In Fig.~\ref{F:pCO_vs_L}(a) the critical values of initial concentration of chunks of knowledge $p_C$ as dependent on linear system size $L$ are presented.
These dependencies become roughly linear in logarithmic scale which indicate that $p_C(L)\propto L^{-\gamma}$.
The least square method fit gives exponent $\gamma=0.4718\ldots$ and $\gamma=0.1895\ldots$ for $K=4$ and $K=8$, respectively.
Of course, for $K=8$ threshold $p_C$ is greater than that for $K=4$, that is, if the organisational change requires more knowledge, more individual portions of knowledge at the beginning of the knowledge transfer (measured by $p$) must be available in the organisation.

Let us look now at the bottom panel of Fig.~\ref{F:omniatT_vs_p}.
As can be seen, for all tested lattice sizes $L$, after reaching the maximum value of $n(K)$ (which takes place at $p=p_O$), the decrease of the fraction $n(K)$ for each test, is observed. 
This decrease takes place until a certain value of $p$ and then $n(K)$ again increases with $p$.
This phenomenon takes place both for $K=4$ and $K=8$ and it is a consequence of the assumptions of our model. 
Agents can acquire knowledge only from neighbours, who have exactly one more of knowledge chunk from them. 
This causes, that at some point agents cannot longer acquire more knowledge, because neighbouring agents are much smarter than them.

The optimal value of initial concentration of chunks of knowledge $p_O$ which guarantee the most efficient transfer of knowledge\footnote{i.e. giving the largest fraction of agents with $K$ chunks of knowledge} as dependent on linear system size $L$ is presented in Fig.~\ref{F:pCO_vs_L}(b).
For $p=p_O$ the dependencies $n(K)$ presented in Fig.~\ref{F:omniatT_vs_p} have local maximum.
In both cases ($K=4$, 8) the dependencies $p_O(L)$ do not grow with $L$.
Please note, that for $K=8$ and small- or average-sized organisations ($L\le 20$) the dependencies $n(K)$ vs. $p$ (see Fig.~\ref{F:omniatT_vs_p}) grow monotonically with $p$ and $p_O=1$.

\subsubsection{\label{S:E4}Fourth experiment: $\langle f\rangle$}

\begin{figure}[!ht]
\centering
\psfrag{p}{$p$}
\psfrag{K=}{$K=$}
\psfrag{L=5}{(a) $L=5$}
\psfrag{L=20}{(b) $L=20$}
\psfrag{L=5, top}{(c) $L=5$}
\psfrag{L=20, top}{(d) $L=20$}
\psfrag{<fci>}[c]{$\langle f\rangle$ [\%]}
\includegraphics[width=.450\textwidth]{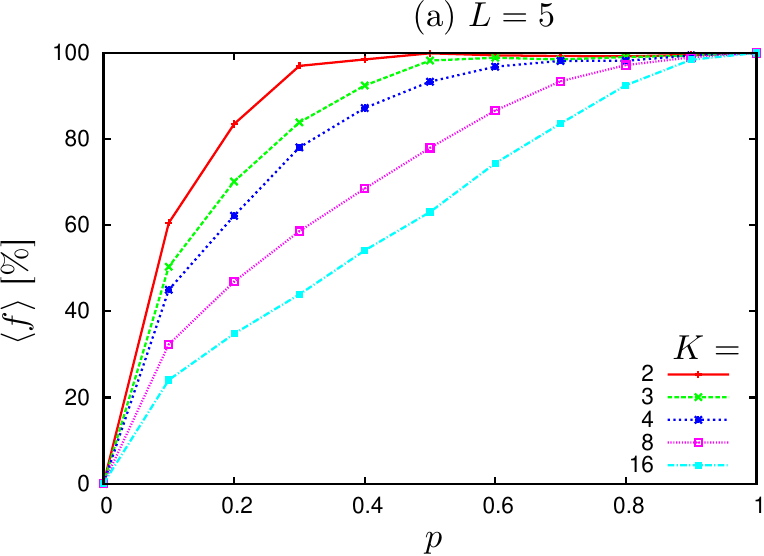}
\includegraphics[width=.450\textwidth]{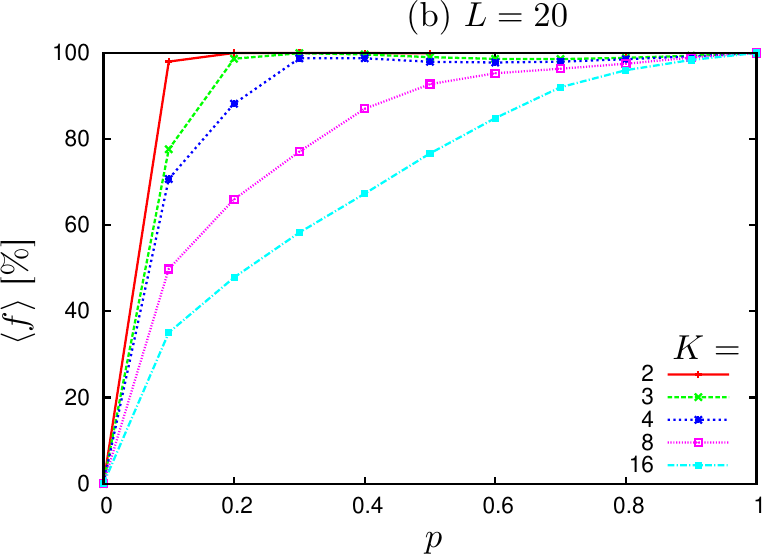}\\
\includegraphics[width=.450\textwidth]{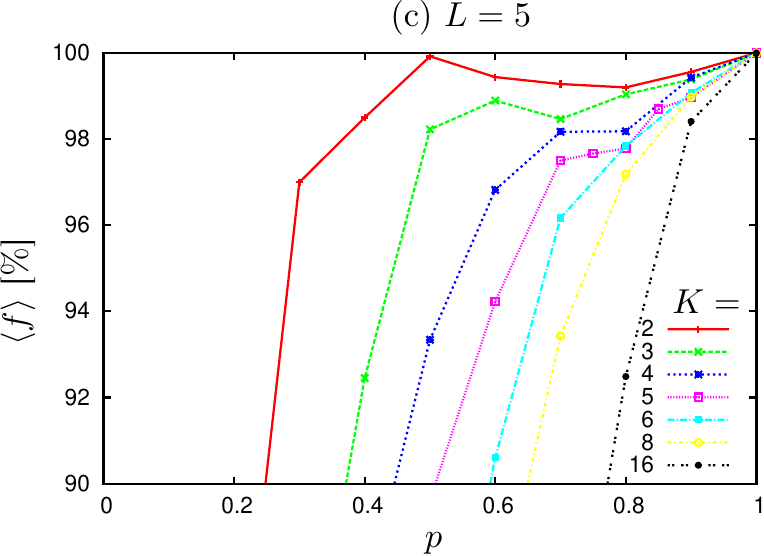}
\includegraphics[width=.450\textwidth]{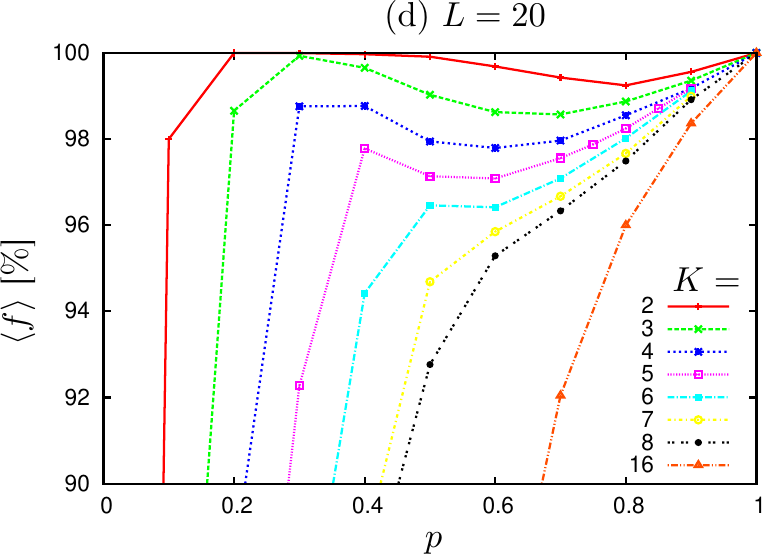}
\caption{\label{F:fci_vs_Kp_v2}The average coverage of chunks of knowledge $\langle f\rangle=K^{-1}\sum_{k=1}^K f(k)$ of agents having any chunk of knowledge $c_k$ for $L=5$, 20 and long simulations time $T\to\infty$ and various values of $K$ ($K=2$, 3, 4, 8, 16) as dependent on initial concentration of chunks of knowledge $p$.
The values of $\langle f\rangle$ are averaged over $M=100$ independent simulations.}
\end{figure}

Subsequently, in the fourth experiment, the effectiveness of knowledge transfer expressed by average coverage of knowledge chunks
\[ \langle f\rangle=\frac{1}{K} \sum_{k=1}^K f(k) \]
of agents having any chunk of knowledge $c_k$ was analysed.
This is a measure which expresses the average number of knowledge chunks possessed by the system (organisation).
Simulations were performed for $K = 2$, 3, 4, 5, 6, 7, 8, 16 and for the $p$ and $L$ levels as in the second experiment. 
In Fig.~\ref{F:fci_vs_Kp_v2} the average coverage of chunks of knowledge $\langle f\rangle$ for $L=5$ and long simulations time $T\to\infty$ as dependent on initial concentration of chunks of knowledge $p$ is presented.
As can be seen in Fig.~\ref{F:fci_vs_Kp_v2}, the greater knowledge $K$ is desired by the organisation, the larger value of $p$ is required to ensure that almost all members of the organisation will receive all $K$ available chunks of knowledge.
For the larger organisation sizes $L$ the smaller values of $p$ are sufficient for observing the effect described above.
The latter is a direct consequence of our model assumptions.
Each agent has independently, with some probability $p$, a required chunk of knowledge. 
Agents (the organisation members) acquire initially knowledge spontaneously and independently.
The probability that the agent will have all of the knowledge is the probabilities product $p^K$ of having each of the knowledge chunks. 
If $p=0.2$, the probability that the agent has, for example, all four required knowledge chunks ($K=4$) is $p^K=0.0016$.
For a small network ($L=5$) initially only a very small number of agents have all the chunks of knowledge $L^2p^K=0.04\ll 1$ and thus we do not observe such agents in lines 5-9 of the Listing~\ref{lst:exmples02}.
For the probability $p$ four times larger, $L^2p^K\approx 10$ and we may expect several agents with $(c_1,c_2,c_3,c_4)=(1,1,1,1)$ at $t=0$.
And indeed, we can detect fourteen such agents present in lines 5-9 of Listing~\ref{lst:exmples08}.
The deviation of observed and expected values of agents with total knowledge lays in both small lattice size ($L=5)$ and extremely low statistical sampling (single realisation, $M=1$).

Let us also see the bottom panel in Fig.~\ref{F:fci_vs_Kp_v2}. 
For $K=2$, 3 and 4  the dependencies $\langle f\rangle$ on $p$ increases to a certain threshold value of $p$ (the threshold value of $p$ is different for various $K$), then it decreases, and it increases again.
These complex behaviours are more visible in the larger lattice ($L=20$) than in the smaller ($L=5$).
We see again, that for $L=20$ and $K\ge 7$ a drop in $\langle f\rangle$ is absent.

\section{\label{S:disc}Discussion and conclusions}

We have investigated a CA model to study transfer of knowledge within the organisation.
The transfer of existing knowledge is studied, because the acquisition of such knowledge is the key to build competitive advantage of companies~\citep{Szulanski-1996}.
Moreover, the knowledge {\em is in} the organisation, as emphasised by \cite{Grant-1996}.
The organisation is considered here in the perspective of complex systems.
In this context, the main role is played by the network of informal contacts \citep{Reagans-2003}.
Our goal was to answer the question: what factors influence the efficiency and effectiveness of knowledge transfer? 
To answer this question, we performed simulations
\begin{itemize}
\item for model organisations of different sizes ($L$),
\item for different sizes of knowledge transfer ($K$),
\item and changing the initial knowledge ($p$), which is in the organisation.
\end{itemize}

As we have shown previously, the larger size of the organisation $L$, the smaller initial fraction $p$ of knowledge chunks among members of the organisation is necessary for reaching (in series of knowledge transfers) the required level of knowledge by almost all members of the organisation.
The value $L^2p^K$ is also an average number of agents $i$ who initially have all chunks of knowledge (i.e. $\nu^i(0)=K$). 
In other words, to obtain finally a similar level of knowledge saturation among agents for larger organisations we need fewer agents with all portions of knowledge before the transfer of knowledge will start when compared to the smaller organisations.
In addition, in the larger organisation it is more likely that the required knowledge will be acquired by almost all the members of the organisation.
This phenomenon results from the assumptions of the  model, as explained above.
The condition for the effective transfer in our model,  is having all knowledge chunks by at least one individual.
If the organisation is small and initial concentration $p$ of chunks of knowledge is low, it may not be the person who has all of the knowledge chunks.
Furthermore, in our model, we assumed that  agents (organisation members) can acquire knowledge only from neighbours, who have exactly one more of knowledge chunk from them.
This assumption stems from the concept of distributed leadership, where agents having the more knowledge are treated as transfer leaders.
As it was shown, this causes, that at some point they can no longer acquire knowledge, because all agents in the neighbourhood  are too much smarter than them.
The adoption of such a way  of knowledge transfer causes that it is sometimes not fully effective.
This is particularly visible in the graphs of Fig.~\ref{F:omniatT_vs_p}(c-d) and Fig.~\ref{F:fci_vs_Kp_v2}.
The radical model assumption that the transfer takes place only between agents, who differ in single chunk of knowledge, may results in the inability to achieve the full knowledge of the organisation.
This phenomenon suggests taking action in the organisation to shorten the distance (social distance) between people with different levels of knowledge (educational attainment), or working out incentives to share knowledge.

\begin{figure}[htbp]
\centering
\psfrag{p}{$p$}
\psfrag{products}[][c]{$\mathcal{B}_2(k_1)\mathcal{B}_2(k_2)$}
\psfrag{01}[l]{$\mathcal{B}_2(0)\mathcal{B}_2(1)$}
\psfrag{02}[l]{$\mathcal{B}_2(0)\mathcal{B}_2(2)$}
\psfrag{12}[l]{$\mathcal{B}_2(1)\mathcal{B}_2(2)$}
\includegraphics[width=.50\textwidth]{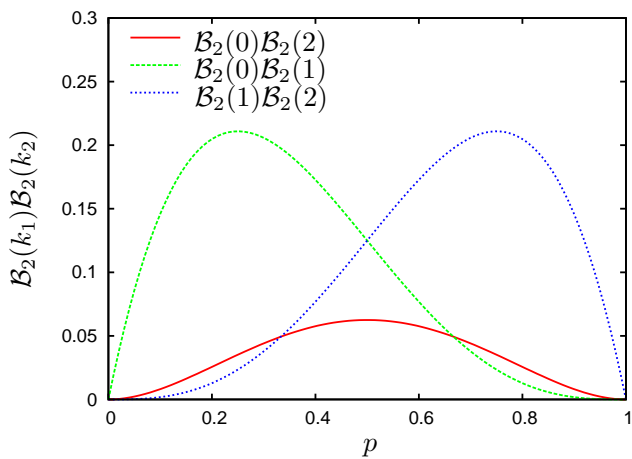}
\caption{\label{F:Bernoulli}The estimated probabilities of acts of knowledge transfer (dotted and dashed lines) and lack of it (solid line) in the first step of simulation as dependent on the initially concentration of chunks of knowledge $p$.}
\end{figure}

The simulation results shown in Fig.~\ref{F:fci_vs_Kp_v2} suggest a decrease in the effectiveness of knowledge transfer with rising probabilities $p$ for the initial individual portions of knowledge.
However, this surprising phenomenon can be explained on the basis of the assumed rules of transfer and assignment of `knowledge' by agents.
Undoubtedly, the dynamics of the knowledge transfer depends on the initial parameters adopted in the model.
The probabilities of initial appearance of agents with $k$ chunks of knowledge are given by Bernoulli distribution
\[
\mathcal{B}_K(k)={K \choose k}p^k(1-p)^{K-k}.
\]
For example, let us assume a simple example of $K=2$---when only two chunks of knowledge are required in the modelled organisation.
The products $\mathcal{B}_2(2)\mathcal{B}_2(1)$, $\mathcal{B}_2(1)\mathcal{B}_2(0)$ and $\mathcal{B}_2(2)\mathcal{B}_2(0)$ provide probabilities of meeting the respective types of agents pairs (and the probabilities of the knowledge transfer or lack of it) for the two neighbouring grid cell $(i,j)$ in the first step of the simulation.
Namely, these products reflect probabilities of appearance of pair of agents with ($k=1$ and $k=2$), ($k=1$ and $k=0$), ($k=2$ and $k=0$) chunks of knowledge, respectively. 
For the first two pairs $\mathcal{B}_2(k_1)\mathcal{B}_2(k_2)$ the transfer of knowledge is possible, while in the third case it is forbidden by the assumed model rules.
These probabilities products are presented in Fig.~\ref{F:Bernoulli}.

For $p\ge 0.5$ we can observe the strong increase of the possibility of the transfer of useful information from agents  with two chunks of knowledge to those who have just one.
Although the number of pairs blocking the knowledge transfer decreases, the drastic decrease of the possibility of obtaining knowledge by agents without any chunks of knowledge from those who have single chunk of knowledge is simultaneously observed.
Hence, in a further steps, the increase of the number of pairs potentially blocking the knowledge transfer may occur and consequently the system does not reach the state of `full knowledge'.
It seems that this reasoning justifies the results shown in Fig.~\ref{F:fci_vs_Kp_v2} for $K = 2$.
Please note however, that our reasoning has rather `mean filed' character and it neglects totally the spatial correlations among agents.
The latter are crucial in CA technique and thus our interpretation remains only qualitative.

Efficiency is understood here as the time $\tau$ needed to achieve a stable state of the system.
The larger the organisation, the time $\tau$ of knowledge transfer is longer.
This time depends also on the number of knowledge chunks $K$.
The more knowledge is needed to be transfer, the time of this process is longer.

As it was postulated by~\cite{Reagans-2003}, effective knowledge transfer is a very important part of the knowledge management.
Without knowledge of the critical factors related to the effective transfer of knowledge, managers may have a problem with supporting the exchange of knowledge~\citep{Levin-2004}.
Our findings could provide the theoretical guidance for organisations to perform the complex knowledge transfer and knowledge management.
Both efficacy and effectiveness studies of knowledge transfer  show a significant role of initial concentration of chunks of knowledge in a transfer process.
If the state of knowledge at the beginning of the transfer is so important, then---in our opinion---organisations should conduct courses and training for their employees.
The more people possess the required knowledge, the more effective and efficient transfer of knowledge will be.
Our proposals concern the spontaneous transfer of knowledge, where the main role is played by informal contacts between employees.
Also, the range of knowledge possessed by organisation members is created spontaneously (random) although it depends on broadly understood intensity of its promotion by the organisation.

\subsection{\label{S:next}Further research}

This is a preliminary study on the transfer of knowledge within the organisation.
The results suggest  further development of the model in the search on factors influencing  more effective and efficient knowledge transfer.

Firstly, we were going to describe the way of knowledge transfer by another or modified rules.
For example, it seems interesting to investigate the transfer of knowledge consisting in acquiring knowledge from individuals who have other pieces of knowledge (not necessarily have more knowledge than a learning agent) and/or acquiring knowledge from the wisest individuals in the neighbourhood (knowledge leaders).
This will allow the comparison of different variants of the model and examine how a different knowledge transfer rules influence  effectiveness  and efficiency of this process.

Secondly, because our research shows a significant role of an initial knowledge in the organisation, it seems to be interesting to study the company's policy involving the different training strategies for their employees.
Undoubtedly managers could be interested in an identification of relationships between the range and scope of training and a quality of knowledge transfer.
For example, the knowledge whether do training for a wider range of employees, or for a smaller number of units; whether to train comprehensively few people, or teach different skills of a larger group of employees, it may prove to be very valuable.

Thirdly, studying the effects of strong and weak ties to the transfer of knowledge, is very interesting~\citep{Uzzi-1997, Uzzi-1999, Hansen-1999}. 
It is related to the introduction of social distance and the different neighbourhood size in our model---here restricted to the smallest possible, i.e. von Neumann neighbourhood where interactions only with the nearest neighbours are considered. 

\subsection*{Acknowledgements}
AKS and KM are grateful to Krzysztof Ku{\l}akowski for fruitful discussion on cellular automaton rules and critical comments on the manuscirpt.
This research was supported by \href{https://www.ncn.gov.pl/?language=en}{National Science Centre} (NCN) in Poland (grant no. UMO-2014/15/B/HS4/04433) and in part by \href{http://www.plgrid.pl/en}{PL-Grid Infrastructure}.

\bibliographystyle{plainnat}
\bibliography{opus8.bib}

\appendix

\section{\label{java}Appendix: Java applet presenting system evolution}

\includegraphics[width=\textwidth]{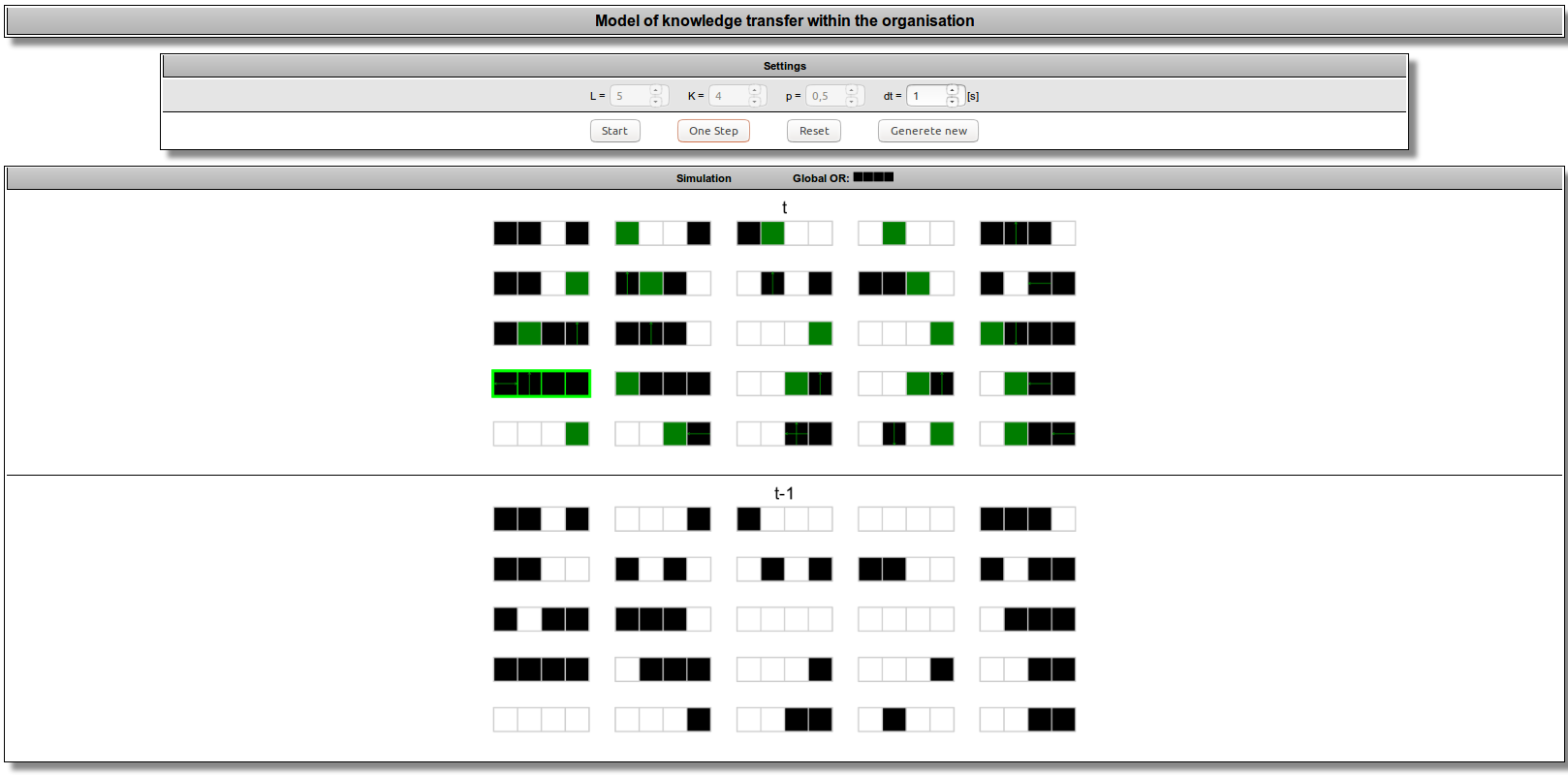}

\url{http://www.zis.agh.edu.pl/knowledge_transfer/}

\section{\label{f77}Appendix: program code listings}

The variables names used in program codes presented in Listings~\ref{lst:main} and~\ref{lst:tau} correspond directly to symbols used in main text body, i.e. p, L, K, T, M, F\_k, N\_k, tau correspond to $p$, $L$, $K$, $T$, $M$, $F(k)$, $N(k)$, $\tau$, respectively.
The two-dimensiomal matrices old and new keep boolean representation of agents chunks of knowledges $c^i_k$ at times $t-1$ and $t$, respectively.
irun counter corresponds to $r$ while L2 stands for $L^2$.
The rand(idum) procedure generates uniformly distributed random numbers from $[0,1)$ interval and idum is a seed of the generator.
The variables avetau and avetau2 allows for estimation of uncertainties $u(\tau)$ [see Eq.~\eqref{eq:u}].
Please manipulate (comment/uncomment) lines 168-179 or 171-182 for pretty printing of program output for $K\ne 4$ in Listings~\ref{lst:main} and~\ref{lst:tau}, respectively.

\lstset{backgroundcolor=\color{lightgray},commentstyle=\itshape\color{blue},stringstyle=\color{teal}}
\begin{lstlisting}[language={[77]Fortran},frame=single,numbers=left,numberstyle=\tiny,basicstyle=\footnotesize,stepnumber=5,caption={Fortan77 code allowing for direct reproduction of the results presented in Figs.~\ref{F:komp_vs_t_v2}-\ref{F:nkomp_vs_t_K8}.},label=lst:main]
      program main
      implicit none 
      real p,rand
      logical old, new, global
      integer L,L2,K,M,T,numchunks,Nglobal,F_k,N_k,
     & i,j,kk,irun,idum,iter,nlockomp,transf
      parameter(L=5,L2=L*L,K=32,M=100,T=200)
      dimension old(-L+1:L2+L,K),new(-L+1:L2+L,K),transf(K),
     & numchunks(-L+1:L2+L),global(K),F_k(K,0:T),N_k(0:K,0:T)
      data idum /1/

      read *, p
      print *,'# L=',L,' K=',K,' M=',M,' p=',p

      do iter=0,T
        N_k(0,iter)=0
        do j=1,K
          F_k(j,iter)=0
          N_k(j,iter)=0
        enddo
      enddo

c ######################################################################
      do 999 irun=1,M

      Nglobal=0
      do j=1,K
        global(j)=.false.
      enddo

      iter=0
C --- initial distribution of competences ---
        do i=1,L2
          numchunks(i)=0
          do j=1,K
            old(i,j)=.false.
            new(i,j)=.false.
            if(rand(idum).lt.p) then
              old(i,j)=.true.
              new(i,j)=.true.
              global(j)=.true.
              numchunks(i)=numchunks(i)+1
            endif
          enddo
        enddo
       
        do j=1,K
          if(global(j)) Nglobal=Nglobal+1
        enddo

        do i=1,L
          numchunks(-L+i)=numchunks(L2-L+i)
          numchunks(L2+i)=numchunks(i)
          do j=1,K
            old(-L+i,j)=old(L2-L+i,j)
            old(L2+i,j)=old(i,j)
          enddo
        enddo

        do i=1,L2
          N_k(numchunks(i),iter)=N_k(numchunks(i),iter)+1
          do j=1,K
            if(new(i,j)) F_k(j,iter)=F_k(j,iter)+1
          enddo
        enddo

      do 998 iter=1,T
C --- one time step of the system evolution ---

        do 777 i=1,L2
        
          if(numchunks(i).eq.numchunks(i-L)-1) then
            nlockomp=0
            do j=1,K
              if((.not.old(i,j)) .and. old(i-L,j)) then
                nlockomp=nlockomp+1
                transf(nlockomp)=j
              endif            
            enddo
            if(nlockomp.gt.0) then
              kk=1+nlockomp*rand(idum)
              new(i,transf(kk))=.true. 
              goto 777
            endif
          endif

          if(numchunks(i).eq.numchunks(i+L)-1) then
            nlockomp=0
            do j=1,K
              if((.not.old(i,j)) .and. old(i+L,j)) then
                nlockomp=nlockomp+1
                transf(nlockomp)=j
              endif
            enddo
            if(nlockomp.gt.0) then
              kk=1+nlockomp*rand(idum)
              new(i,transf(kk))=.true.
              goto 777
            endif
          endif

          if(numchunks(i).eq.numchunks(i-1)-1) then
            nlockomp=0
            do j=1,K
              if((.not.old(i,j)) .and. old(i-1,j)) then
                nlockomp=nlockomp+1
                transf(nlockomp)=j
              endif
            enddo
            if(nlockomp.gt.0) then
              kk=1+nlockomp*rand(idum)
              new(i,transf(kk))=.true.
              goto 777
            endif
          endif

          if(numchunks(i).eq.numchunks(i+1)-1) then
            nlockomp=0
            do j=1,K
              if((.not.old(i,j)) .and. old(i+1,j)) then
                nlockomp=nlockomp+1
                transf(nlockomp)=j
              endif
            enddo
            if(nlockomp.gt.0) then
              kk=1+nlockomp*rand(idum)
              new(i,transf(kk))=.true.
              goto 777
            endif
          endif

 777    enddo

C --- synchronous system update ---
        do i=1,L2
          numchunks(i)=0
          do j=1,K
            old(i,j)=new(i,j)
            if(old(i,j)) numchunks(i)=numchunks(i)+1
          enddo
        enddo

        do i=1,L
          numchunks(-L+i)=numchunks(L2-L+i)
          numchunks(L2+i)=numchunks(i)
          do j=1,K
            old(-L+i,j)=old(L2-L+i,j)
            old(L2+i,j)=old(i,j)
          enddo
        enddo

        do i=1,L2
          N_k(numchunks(i),iter)=N_k(numchunks(i),iter)+1
          do j=1,K
            if(new(i,j)) F_k(j,iter)=F_k(j,iter)+1
          enddo
        enddo

 998  enddo
      
 999  enddo

      print *,""
      print '(a6,a66,a72)','#it |',
     & 'F(c_1), ..., F(c_K) |',
     & 'N(0), ..., N(K) -- data aggregated from M runings'
      do iter=0,T
c       K=32:
c       print '(i4,a2,32i8,a2,33i8)',iter,
c       K=16:
c       print '(i4,a2,16i8,a2,17i8)',iter,
c       K=8:
c       print '(i4,a2,8i8,a2,9i8)',iter,
c       K=4:
        print '(i4,a2,4i8,a2,5i8)',iter,
c       K=3:
c       print '(i4,a2,3i8,a2,4i8)',iter,
c       K=2:
c       print '(i4,a2,2i8,a2,3i8)',iter,
     &        '|',(F_k(j,iter),j=1,K),
     &        '|',(N_k(j,iter),j=0,K)
      enddo

      end
\end{lstlisting}

\begin{lstlisting}[language={[77]Fortran},frame=single,numbers=left,numberstyle=\tiny,basicstyle=\footnotesize,stepnumber=5,caption={Fortan77 code allowing for reproduction of data presented in Fig.~\ref{F:tau_vs_p}.},label=lst:tau]
      program tau
      implicit none 
      real p,rand
      logical old, new, global, change
      integer L,L2,K,M,T,numchunks,Nglobal,F_k,N_k,F_k_old,N_k_old,
     & i,j,kk,irun,idum,iter,nlockomp,transf,tau
      real avetau, avetau2
      parameter(L=20,L2=L*L,K=4,M=100)
      dimension old(-L+1:L2+L,K),new(-L+1:L2+L,K),transf(K),
     & numchunks(-L+1:L2+L),global(K),F_k(K),N_k(0:K),
     & F_k_old(K),N_k_old(0:K)
      data idum,avetau,avetau2 /1, 0.0, 0.0/

      read *, p
      print *,'# L=',L,' K=',K,' M=',M,' p=',p

c ######################################################################
      do 999 irun=1,M
      print *,'# irun=',irun

      print *,""
      print '(a6,a66,a72)','#it |',
     & 'F(c_1),  ..., F(c_K) |',
     & 'N(0), ..., N(K)'

      N_k_old(0)=0
      do j=1,K
        F_k_old(j)=0
        N_k_old(j)=0
      enddo

      Nglobal=0
      do j=1,K
        global(j)=.false.
      enddo

      iter=0
C --- initial distribution of competences ---
      do i=1,L2
        numchunks(i)=0
        do j=1,K
          old(i,j)=.false.
          new(i,j)=.false.
          if(rand(idum).lt.p) then
            old(i,j)=.true.
            new(i,j)=.true.
            global(j)=.true.
            numchunks(i)=numchunks(i)+1
          endif
        enddo
      enddo
       
      do j=1,K
        if(global(j)) Nglobal=Nglobal+1
      enddo

        do i=1,L
          numchunks(-L+i)=numchunks(L2-L+i)
          numchunks(L2+i)=numchunks(i)
          do j=1,K
            old(-L+i,j)=old(L2-L+i,j)
            old(L2+i,j)=old(i,j)
          enddo
        enddo

        do i=1,L2
          N_k(numchunks(i))=N_k(numchunks(i))+1
          do j=1,K
            if(new(i,j)) F_k(j)=F_k(j)+1
          enddo
        enddo

 998  iter=iter+1
C --- one time step of the system evolution ---

        N_k(0)=0
        do j=1,K
          F_k(j)=0
          N_k(j)=0
        enddo

        do 777 i=1,L2
        
          if(numchunks(i).eq.numchunks(i-L)-1) then
            nlockomp=0
            do j=1,K
              if((.not.old(i,j)) .and. old(i-L,j)) then
                nlockomp=nlockomp+1
                transf(nlockomp)=j
              endif            
            enddo
            if(nlockomp.gt.0) then
              kk=1+nlockomp*rand(idum)
              new(i,transf(kk))=.true. 
              goto 777
            endif
          endif

          if(numchunks(i).eq.numchunks(i+L)-1) then
            nlockomp=0
            do j=1,K
              if((.not.old(i,j)) .and. old(i+L,j)) then
                nlockomp=nlockomp+1
                transf(nlockomp)=j
              endif
            enddo
            if(nlockomp.gt.0) then
              kk=1+nlockomp*rand(idum)
              new(i,transf(kk))=.true.
              goto 777
            endif
          endif

          if(numchunks(i).eq.numchunks(i-1)-1) then
            nlockomp=0
            do j=1,K
              if((.not.old(i,j)) .and. old(i-1,j)) then
                nlockomp=nlockomp+1
                transf(nlockomp)=j
              endif
            enddo
            if(nlockomp.gt.0) then
              kk=1+nlockomp*rand(idum)
              new(i,transf(kk))=.true.
              goto 777
            endif
          endif

          if(numchunks(i).eq.numchunks(i+1)-1) then
            nlockomp=0
            do j=1,K
              if((.not.old(i,j)) .and. old(i+1,j)) then
                nlockomp=nlockomp+1
                transf(nlockomp)=j
              endif
            enddo
            if(nlockomp.gt.0) then
              kk=1+nlockomp*rand(idum)
              new(i,transf(kk))=.true.
              goto 777
            endif
          endif

 777    enddo

C --- synchronous system update ---
        do i=1,L2
          numchunks(i)=0
          do j=1,K
            old(i,j)=new(i,j)
            if(old(i,j)) numchunks(i)=numchunks(i)+1
          enddo
        enddo

        do i=1,L
          numchunks(-L+i)=numchunks(L2-L+i)
          numchunks(L2+i)=numchunks(i)
          do j=1,K
            old(-L+i,j)=old(L2-L+i,j)
            old(L2+i,j)=old(i,j)
          enddo
        enddo

        do i=1,L2
          N_k(numchunks(i))=N_k(numchunks(i))+1
          do j=1,K
            if(new(i,j)) F_k(j)=F_k(j)+1
          enddo
        enddo

c       K=32:
c       print '(i4,a2,32i8,a2,33i8)',iter,
c       K=16:
c       print '(i4,a2,16i8,a2,17i8)',iter,
c       K=8:
c       print '(i4,a2,8i8,a2,9i8)',iter,
c       K=4:
        print '(i4,a2,4i8,a2,5i8)',iter,
c       K=3:
c       print '(i4,a2,3i8,a2,4i8)',iter,
c       K=2:
c       print '(i4,a2,2i8,a2,3i8)',iter,
     &        '|',(F_k(j),j=1,K),
     &        '|',(N_k(j),j=0,K)

      change=.false.

      do j=0,K
        if(N_k(j) .ne. N_k_old(j)) change=.true.
      enddo

      do j=1,K
        if(F_k(j) .ne. F_k_old(j)) change=.true.
      enddo

      if(change) then
        N_k_old(0)=N_k(0)
        do j=1,K
          F_k_old(j)=F_k(j)
          N_k_old(j)=N_k(j)
        enddo
        goto 998
      else 
        tau=iter
        avetau=avetau+1.0*tau
        avetau2=avetau2+1.0*tau*tau
        print *,tau
        goto 999
      endif

 999  enddo

      avetau=avetau/(1.0*M)
      avetau2=avetau2/(1.0*M)
      print *, p, avetau, sqrt((avetau2-avetau*avetau)/(1.0*M-1.0))
      end
\end{lstlisting}
\end{document}